\documentclass[preprint,aps,showpacs]{revtex4-1}
\usepackage{graphicx,epsfig}% Include figure files
\usepackage{dcolumn}% Align table columns on decimal point
\usepackage{bm}% bold math
\usepackage[dvipsnames,usenames]{xcolor}
\usepackage{amsmath}
\usepackage{amssymb} % AMS
\usepackage{hyperref}
\usepackage{xcolor}
\usepackage{pgfplots}
\usepackage{bbold}
\usepackage{float}

\pgfplotsset{compat=1.5}

\begin{document}
\preprint{LA-UR-19-29015}
\title{Quasielastic lepton scattering and back-to-back nucleons in
the short-time approximation}
\author{ S.~Pastore}
\email{saori@wustl.edu}
\affiliation{Physics Department and McDonnell Center for the Space Sciences at
Washington University in St.~Louis, MO, 63130, USA}
\author{ J.~Carlson}
\email{carlson@lanl.gov}
\author{ S.~Gandolfi}
\email{stefano@lanl.gov}
\affiliation{Theoretical Division, Los Alamos National Laboratory, Los Alamos, NM, 87545, USA}
\author{R.~Schiavilla}
\email{schiavil@jlab.org}
\affiliation{ Department of Physics, Old Dominion University, Norfolk, VA 23529, USA\\
 Jefferson Lab, Newport News, VA 23606, USA}
\author{ R.B.~Wiringa}
\email{wiringa@anl.gov}
\affiliation{Physics Division, Argonne National Laboratory, Argonne, IL 60439, USA}
\date{\today}
\begin{abstract}
Understanding quasielastic electron- and neutrino-scattering from nuclei has taken on
new urgency with current and planned neutrino oscillation experiments, and with electron
scattering experiments measuring specific final states, such as those involving nucleon
pairs in ``back-to-back'' configurations.  Accurate many-body methods are available for
calculating the response of light ($A \leq 12$) nuclei to electromagnetic and weak probes,
but they are computationally intensive and only applicable to the inclusive response.  In
the present work we introduce a novel approach, based on realistic models of nuclear
interactions and currents, to evaluate the short-time (high-energy) inclusive and exclusive
response of nuclei.  The approach accounts reliably for crucial two-nucleon dynamics,
including correlations and currents, and provides information on back-to-back nucleons
observed in electron and neutrino scattering experiments.  We demonstrate that in the
quasielastic regime and at moderate momentum transfers both initial- and final-state
correlations, and two-nucleon currents are important for a quantitatively successful
description of the inclusive response and final state nucleons.  Finally, the approach can
be extended to include relativistic---kinematical and dynamical---effects, at
least approximately in the two-nucleon sector, and to describe the response in the resonance-excitation
region.
\end{abstract}
\maketitle
\section{Introduction}

Lately, there has been a resurgence of  interest in quasielastic scattering of electrons
and neutrinos off nuclei, mostly driven by the increasing relevance of accelerator neutrino 
experiments in determining fundamental properties of these particles, such as the oscillation
parameters and charge-conjugation
and parity violating phase~\cite{t2k_web,hk_web,dune_web,mb_web}, and by the many
recent electron scattering experiments finding a significant fraction of events
with back-to-back nucleons, primarily neutron-proton pairs, in the final
state~\cite{JLAB,JLABHALLA,Subedi:2008zz,Hen614}.

While seemingly a simple process, inclusive quasielastic scattering at moderate momentum
transfers is in fact more subtle than originally thought.  The cleanest experimental evidence for this
is the Rosenbluth separation of the inclusive $(e,e^\prime)$ cross-section~\cite{RevModPhys.80.189,Carlson:2001mp},
where the longitudinal and transverse response functions differ by approximately 30\%, after dividing
out the relevant electric and magnetic nucleon form factors. The excess of transverse over
longitudinal strength, which exists on both the low- and high-energy side of the quasielastic peak
for moderate momentum transfers ($2$--$4$ fm$^{-1})$, is caused predominantly by two-nucleon
processes, both initial and final state correlations and two-body currents.  The importance of
these two-nucleon processes has been known for quite some
time~\cite{Carlson:1994zz,Carlson:1997,Bacca:2014tla,VanderSluys:1995rp,Bauer:2000cz,Carlson:2001mp,Amaro:2010sd,Martini:2011wp}
and has most recently been confirmed by accurate Quantum Monte Carlo (QMC) calculations of electron 
scattering off $^{12}$C~\cite{Lovato:2013,Lovato:2015qka,Lovato:2016gkq}.  

In this paper we present simple arguments that show how the quasielastic response arises
primarily from the short-time response of one and two nucleons, and how this fact leads to
the observed scaling in momentum transfer ($y$-scaling) and nuclear mass
(superscaling)~\cite{Donnelly:1999sw}.  We also introduce an approach to calculate the short-time
propagation resulting from two-nucleon dynamics.  The approach yields results in agreement with
those of the more accurate Green's Function Monte Carlo (GFMC) method for inclusive response.
It also provides information on the energy distribution of the interacting pair right after the electroweak
interaction has occurred (i.e., at the vertex), and hence, at least for light nuclei, on exclusive channels involving two nucleons
in the final state.  In heavier nuclei, the additional interactions between the pair
and spectator nucleons needed for a more reliable treatment of these exclusive channels can be approximated by semi-classical event generators. 

The approach we propose only requires knowledge of the ground-state, and thus can be used to study
heavy nuclei amenable to auxiliary-field diffusion Monte Carlo (AFDMC) calculations~\cite{Carlson:2014vla}.
Further, since it involves only two active nucleons, it can be improved to include relativistic (kinematical
and dynamical) effects and pion production channels. 

The paper is structured as follows.  In Secs.~\ref{sec:nuclearRes} and~\ref{sec:twobodyphysics} we
discuss the role of two-body physics---two-nucleon correlations and currents---in 
electromagnetic longitudinal and transverse response functions, and in Sec.~\ref{sec:scaling} the
QMC approach, based on the imaginary-time formalism, is used to calculate {\it ab initio}
these response functions.  In Sec.~\ref{sec:realtime} we introduce the Short-Time Approximation (STA),
which in essence accounts for the full propagation of
nucleon pairs in real rather than imaginary time.  
Results in the STA for inclusive scattering and some exclusive channels, specifically
those involving nucleon pairs in back-to-back kinematics, are given in Secs.~\ref{sec:inclusiveresults} and~\ref{sec:exclusiveresults},
respectively.   Final remarks and conclusions are summarized in Sec.~\ref{sec:summary}.
We relegate to Appendix~\ref{app:inter} details on the two-body dynamics 
leading to the observed excess of strength in the transverse response.

\section{Nuclear Response Functions}
\label{sec:nuclearRes}

The quasielastic inclusive-scattering cross section of electrons and neutrinos by nuclei is written
in terms of electroweak response functions, see Refs.~\cite{Schiavilla:1989jm,Carlson:2001mp,Shen:2012xz} 
for the complete expressions.  For the electromagnetic case of primary interest in the present work,
there are two response functions, namely a longitudinal and a transverse one, schematically given by 
\begin{equation}	
       R_\alpha (q,\omega) = 
	{\overline{\sum_{M_i}}} \sum_f   \langle \Psi_i | O_\alpha ^\dagger ({\bf q}) | \Psi_f \rangle	 
		 \langle \Psi_f |  O_\alpha ({\bf q}) | \Psi_i \rangle \, \delta(E_f - E_i - \omega)\ ,
\end{equation}
where $O_\alpha({\bf q})$ is the electromagnetic charge ($\alpha\,$=$\, L$) or current ($\alpha\,$=$\, T$)
operator.  Here, $\omega$ and ${\bf q}$ 
are the energy and three-momentum transferred to the nucleus, $| \Psi_i \rangle$ and $| \Psi_f \rangle$
represent, respectively, the initial ground state and final continuum state with energies
$E_i$ and $E_f$, and an average over the initial spin projections $M_i$ of the initial state with spin $J_i$
(indicated by the overline) is implied.

The response can be equivalently written as the
matrix element of a current-current correlator by replacing the sum over final states
with a real-time propagator, namely
\begin{equation}	
R_\alpha (q,\omega) = 
  \int_{-\infty}^\infty  \frac{d t}{2 \pi} \,
  {\rm e}^{ i \left(\omega+E_i\right)   t }\, \overline{\sum_{M_i}} \,\langle \Psi_i | 
 	  O_\alpha^\dagger ({\bf q})\,{\rm e}^{-i H t}\,
 	  O_\alpha ({\bf q}) |\Psi_i \rangle \ .
\label{eq:realtime}
\end{equation}
In the equation above, the many-body nuclear Hamiltonian is taken to consist
of single-nucleon kinetic energy terms, and two- and three-nucleon interactions
\begin{equation}
H = \sum_i -\frac{\hbar^2}{2m} \,{\bm \nabla}_i^2+ \sum_{i<j} v_{ij} + \sum_{i<j<k} V_{ijk} \ .
\end{equation}
The charge and current operators are also written as sums of one- and two-nucleon terms (and, in principle, 
many-nucleon terms, though they are ignored in the present work),
\begin{equation}
O_\alpha({\bf q})= \sum_i  O^{(\alpha)}_i({\bf q}) + \sum_{i<j}  O^{(\alpha)}_{ij} ({\bf q})+ \cdots \ .
\end{equation}
The nucleon and nucleon-to-$\Delta$ transition electromagnetic form factors
entering these charge and current operators $O^{(\alpha)}_i({\bf q})$ and
$O^{(\alpha)}_{ij}({\bf q})$ use
standard parametrizations---dipole for the proton electric and magnetic, and neutron
magnetic form factors, and the Galster form for the
neutron electric form factor, see, for example, Ref.~\cite{Shen:2012xz}---and are evaluated at the
four-momentum transfer $Q^2_{\rm qe}=q^2-\omega_{\rm qe}^2$ with
$\omega_{\rm qe}=\sqrt{q^2+m^2}-m$, where $m$ is the nucleon mass.
Other parametrizations or calculations of the nucleon form factors,
for example the $z$-expansion~\cite{Meyer:2016oeg} or calculations from lattice gauge 
theory~\cite{Jang:2019jkn,Rajan:2017lxk,Ishikawa:2018rew,Shintani:2018ozy,Alexandrou:2017ypw,Hasan:2019noy,Alexandrou:2018sjm}  can be easily included.
Two-nucleon terms in both the interactions and currents---collectively 
indicated by ``two-body physics''---are dominated by one-pion-exchange dynamics.

It is also useful to consider sum rules associated with these response
functions, as they provide an indication of the overall contribution from two-nucleon currents,
\begin{equation}
\label{eq:sumrule}
G^2_\alpha(Q^2_{\rm qe})\, S_\alpha(q) = \int^\infty_{\omega_{\rm el}} {d \omega}\, R_\alpha(q,\omega) 
= \overline{\sum_{M_i}} \langle \Psi_i | 
 	  O_\alpha^\dagger ({\bf q})\, O_\alpha ({\bf q}) |\Psi_i \rangle  \ ,
% 	  - |\langle \Psi_i| {\cal O} ({\bf q}) |\Psi_i \rangle |^2\ ,
\end{equation}
where $\omega_{\rm el}\,$=$\,\sqrt{q^2+m_i^2}-m_i$ is the threshold for elastic scattering ($m_i$ is
the rest mass of the initial nucleus).  Note that we calculate the sum rule
corresponding to the response of point-like nucleons, hence the factor $G^2_\alpha(Q^2_{\rm qe})$
denoting the square of the appropriate combination of nucleon electromagnetic
form factors~\cite{Carlson:1997,Lovato:2014} is removed from the sum rule.  The
definition above includes the elastic contribution; the inelastic sum rule is obtained as
\begin{equation}
S^{\rm inel}_\alpha(q)=S_\alpha (q) -\overline{\sum_{M_i}} \sum_{M_{i^\prime}}
\frac{|\langle \Psi_{i^\prime} | O_\alpha({\bf q}) |\Psi_i \rangle |^2}{G^2_\alpha(Q^2_{\rm el})} \ ,
\end{equation}
where the last term in the equation above is the elastic form factor and the
nucleon form factor combination is now evaluated at $Q_{\rm el}^2\,$=$\,q^2-\omega_{\rm el}^2$.
Note that for (initial) nuclear states
with $J^\pi=0^+$, such as $^4$He, there is no elastic contribution associated with magnetic transitions
(namely, for $\alpha\,$=$\,T$).

\begin{figure}[t]
    \centering
	\includegraphics[height=3.0in]{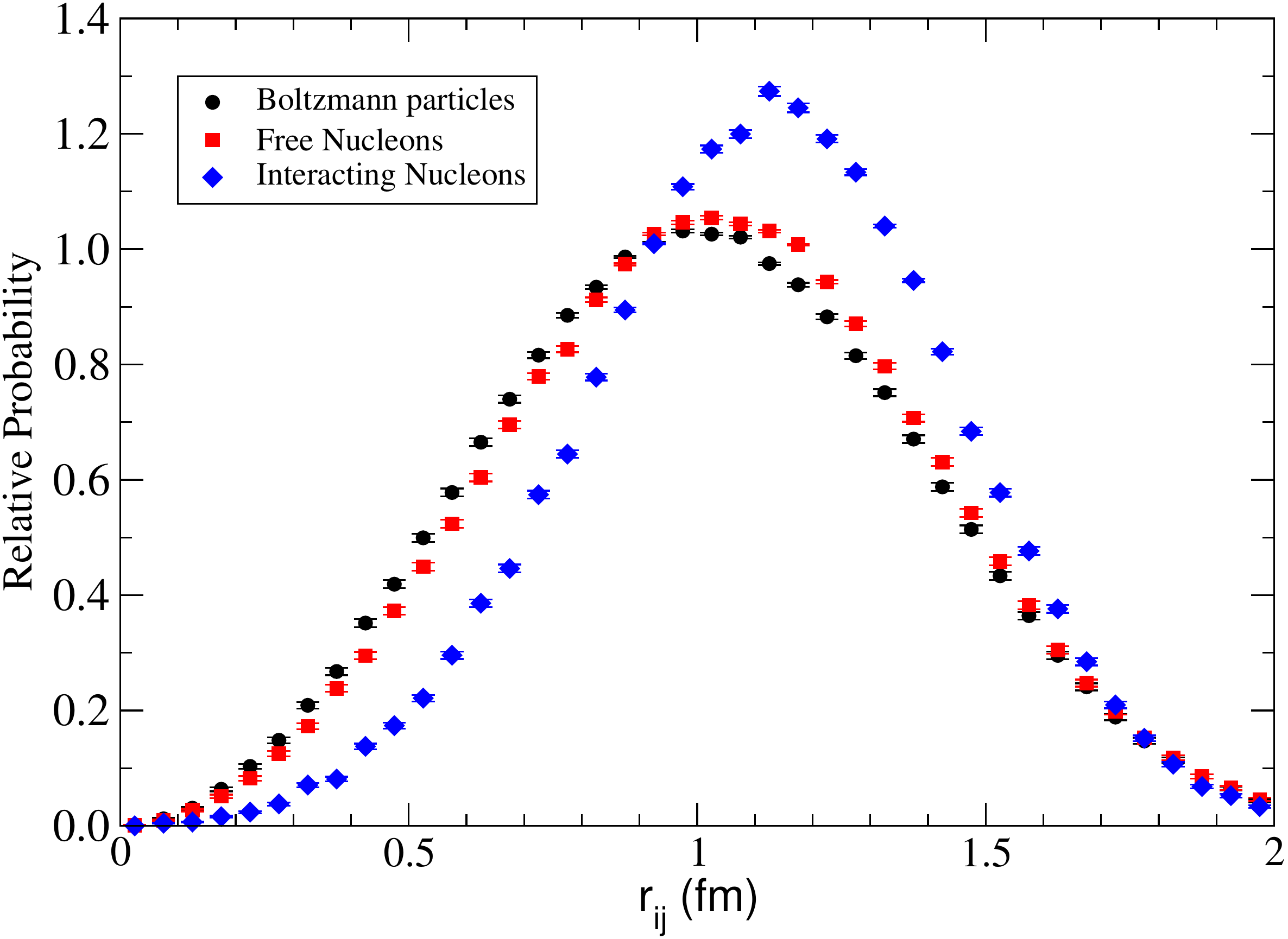}
    \caption{ 
	Nearest-neighbor probability density at nuclear matter density for non-interacting Boltzmann (distinguishable) particles (black circles),
	free nucleons (red squares) and with realistic
	interactions (blud diamonds).
    }
    \label{fig:neighbors}
\end{figure}

It is not surprising that two-nucleon processes of one-pion-range play an important role in quasi-elastic scattering at moderate momentum transfers.
The central density of atomic nuclei is $\sim\,$0.16 fm$^{-3}$, corresponding
to a Fermi momentum $k_F$ of $\sim\,$$1.35$ fm$^{-1}$ or $270$
MeV/c.  A simple cubic solid at $\rho\,$=$\, 0.16$ fm$^{-3}$ would have a nearest-neighbor distance or lattice spacing of about 1.9 fm.
A liquid will have fluctuations that produce, on average, smaller
nearest neighbor distances.  
In Fig.~\ref{fig:neighbors}  we plot the distribution of nearest-neighbor
distances for free Boltzman (distinguishable) particles at nuclear matter density.  A simple density response for this system would be fully incoherent.
We also plot the nearest-neighbor distance distributions for free and
interacting nucleons at the same density. These distributions are obtained by
sampling from the square of the wave function summed over spin-isospin states
to obtain samples of the 3A coordinates, and then for each nucleon
finding the nearest neighboring nucleon.

These distributions all peak at around
$1.1$ fm, very similar to the range of the one-pion-exchange (OPE) interaction.
The interacting distribution is smaller at very short distances but larger
near the peak, reflecting the repulsion and attractions at very short and
moderate distances.
These considerations also fit with the picture that has emerged from {\it ab initio}
studies of nuclear structure~\cite{Forest:1996kp}, that the two-nucleon probability density as function
of the relative separation $r_{ij}$ peaks at about 1 fm
for nucleon pairs in spin/isospin states $S/T\,$=0/1 (quasi-bound $^1$S$_0$ channel) and
1/0 (deuteron-like channel), in which the OPE interaction
plays a major role.

The relevant relative pair momentum corresponding to these inter-nucleon separations
is $\pi/r_{ij}\,$$\sim$ 500 MeV/c.   Only at momenta much higher than this value, when
the corresponding nearest-neighbor probability density is much reduced, can the scattering
be regarded as entirely incoherent, and many-body effects can be neglected.

\section{Two-Nucleon Currents and Correlations}
\label{sec:twobodyphysics}

% 
% 
% \saori{to Rocco: could you please add some rpa references? some old
% stuff?}

It is important to understand how two-nucleon processes enter the quasi-elastic
response, and in particular how they affect its energy dependence.  They have
been found to give $~30\%$ contributions to the electromagnetic transverse
response~\cite{Carlson:2001mp,Lovato:2015qka} and also to contribute substantially to the axial
transverse response~\cite{Lovato:2017cux,Martini:2011wp,Martini:2013sha,Nieves:2013fr}.  The calculations
of Refs.~\cite{Carlson:2001mp,Lovato:2015qka,Lovato:2017cux} show that this enhancement
comes about because of constructive interference between the matrix elements of
the one-body current and the (leading) two-body current induced by pion exchange~\cite{Carlson:2001mp,Lovato:2015qka}.
In this connection, it should be stressed that the contributions of these two-nucleon currents
would be largely suppressed if correlations in the nuclear wave functions were to be
turned off~\cite{Carlson:2001mp}.  Thus, this large excess of 
transverse strength results from the interplay between two-nucleon correlations and 
two-nucleon currents, both induced primarily by OPE
dynamics.  The discussion to follow is meant to illustrate these two aspects---the role
of correlations and the constructive interference between one- and two-body matrix
elements---and, in particular, their complementarity.

\begin{figure}[htbp]
\centering
\includegraphics[width=0.7\linewidth]{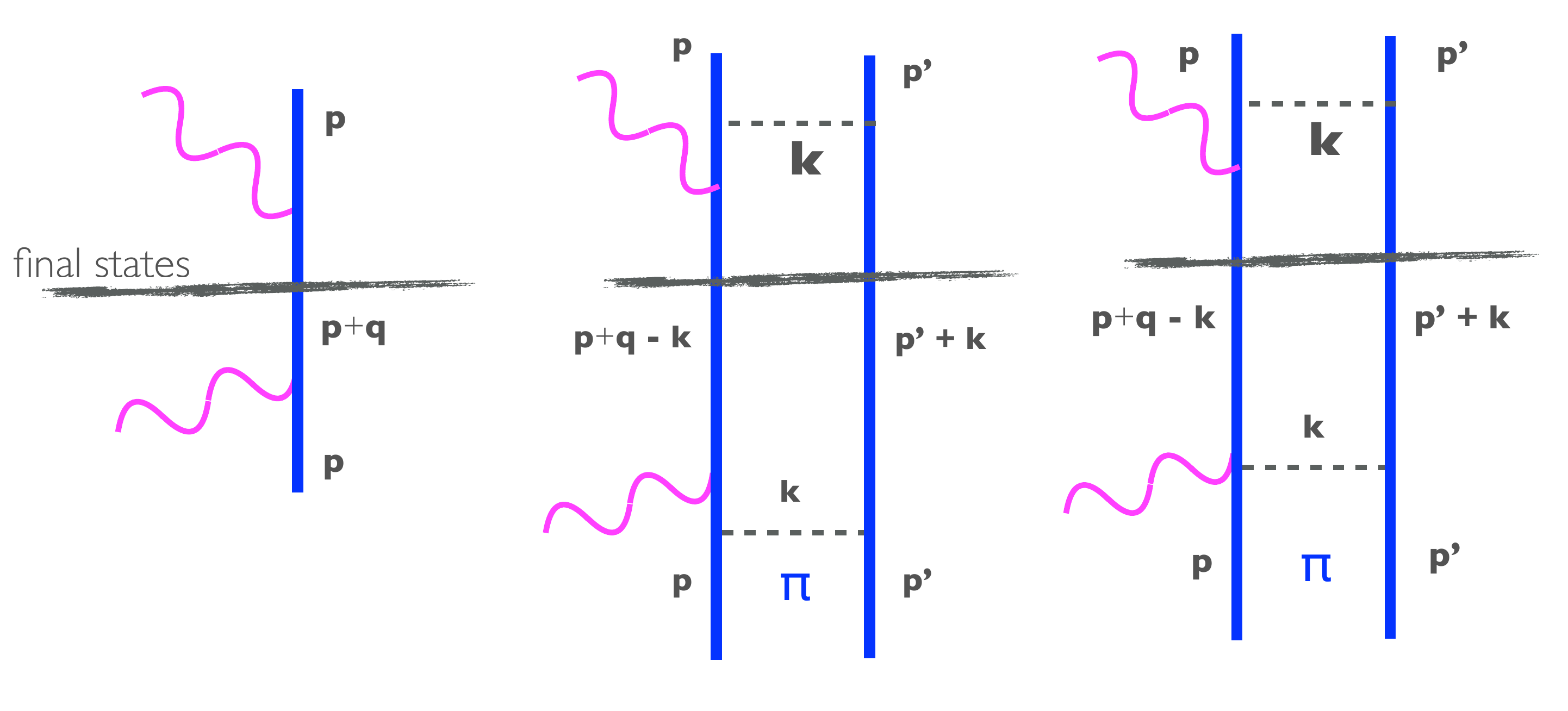}
\caption{Contributions to the sum rule from single-nucleon currents in the plane-wave
	impulse approximation (PWIA) from low- (left panel) and higher-momentum nucleons
	(middle panel). The latter interfere constructively with the two-nucleon current
	operator as indicated in the right panel. See text for further explanations.}
\label{fig:diagrams}
\end{figure}

The simplest way to elucidate these features is to consider one- and two-nucleon
contributions to the (transverse) sum rule defined in Eq.~(\ref{eq:sumrule}) (these
are in fact the largest contributions by far~\cite{Carlson:2001mp}),
\begin{equation}
\langle \Psi_i| O^\dagger \, O|\Psi_i\rangle \simeq 
\langle \Psi_i| \sum_i O_i^\dagger \, O_i+\sum_{i\ne j}O_i^\dagger \, O_j
+\sum_{i<j}\left[ \left(O_i+O_j\right)^\dagger\, O_{ij}+{\rm h.c.}\right] +
\sum_{i<j} O_{ij}^\dagger O_{ij} |\Psi_i\rangle  \ .
\end{equation}
We insert in the expression above complete sets of two-nucleon states $|\psi_{ij}\rangle$, which
satisfy the Lippmann-Schwinger equation
\begin{eqnarray}
\label{eq:Lippmann}
 |\psi_{ij}\rangle &=& |\phi_{ij}\rangle + \frac{1}{e-H^0_{ij}+i\eta}\, v_{ij}^\pi \,|\psi_{ij}\rangle   \nonumber \\
              &=& |\phi_{ij}\rangle + \frac{1}{e-H^0_{ij}+i\eta}\, v_{ij}^\pi\,  |\phi_{ij}\rangle + \cdots\,\,
              \simeq\,\, |\phi_{ij}\rangle + \frac{v_{ij}^{\pi}}{{e}(k)}\, |\phi_{ij}\rangle\ , 
\end{eqnarray}
where in the last step we have treated the OPE interaction $v_{ij}^\pi$ as a perturbation to the free
two-nucleon Hamiltonian $H_{ij}^0$.  Here $|\phi_{ij}\rangle$ represents the (two-nucleon)
free state and $H^{ij}_0\, |\phi_{ij}\rangle$=$\, e |\phi_{ij}\rangle$, where $e(k)$ is
the overall energy denominator associated with the final state.  This energy denominator
depends on the exchanged momentum $k$, as indicated in Fig.~\ref{fig:diagrams}.
The insertion of these complete sets of two-nucleon states is illustrated schematically
by the second and third diagrams of Fig.~\ref{fig:diagrams}. In particular, OPE correlations
are represented by a pion (dashed line) being exchanged between two nucleons (solid lines).
The one-body current operators are indicated by the vertex involving a photon (wavy line) 
interacting with a single nucleon, and two-body current operators are indicated
by a photon interacting with a nucleon and producing a pion, which
is then reabsorbed by the other nucleon.  This so-called seagull term is known
to be dominant among two-body currents of OPE range
(for instance, contributions associated with pion-in-flight currents
are generally found to be suppressed relative to those induced by
seagull currents~\cite{Carlson:1997,Bacca:2014tla}).
In the present qualitative discussion, we are only accounting for
leading terms in both the two-nucleon correlations
and two-nucleon currents. 

We can now analyze the one- and two-nucleon terms of the sum rules illustrated in 
Fig.~\ref{fig:diagrams} and  consider the contribution of particular intermediate 
energy states right after the electromagnetic vertex.
At a modestly large momentum transfer scale $q \sim 500$ MeV/c, considerably larger
than the Fermi momentum of nuclear matter at equilibrium, 
Pauli blocking is not important, and the ground-state can be thought of as consisting 
of a large component of low-momentum nucleons (those described by a mean-field or simple 
shell-model picture), with a significant fraction of high-momentum nucleons predominantly 
produced by OPE correlations.
The incoherent terms in the cross section are then dominated by the left and middle panels
in Fig.~\ref{fig:diagrams}. The two-body final states are to be summed over as indicated by
the cut (horizontal line) in the figure. The left diagram describes the contribution of the
low-momentum mean-field nucleons. The high-momentum contribution shown in the middle
diagram spreads the response over a wider region of energy since the initial momentum 
is uncorrelated (it can be parallel, orthogonal or anti-parallel) with the momentum transfer ${\bf q}$.
 
In addition, though, the high-momentum final states can interfere with two-nucleon
currents that also produce high-momentum final states. This process is shown in the
right diagram. There are two different orderings in the squared matrix element, only
one of which is shown---the seagull current. (We emphasize
that the results presented in later sections retain all two-body terms, including
those associated with $\Delta$ excitation.)
This particular process---also referred to as the ``one- and two-body interference term''---has been 
determined to be quite important in quasi-elastic  scattering~\cite{Carlson:1997,Carlson:2001mp}.
The kinematics of the second and third diagrams are very similar,
they lead to fairly high relative-momentum final states, but the strength of the response
in total energy is spread across the quasielastic peak. Electron scattering experiments 
have been performed to understand these high-momentum nucleons in more detail, finding 
roughly 20\% probability for nucleons to be above the Fermi momentum $k_F$~\cite{Subedi:2008zz,Hen614,Arrington:2011xs}.  

Let us now turn our attention to the mechanism leading to the excess of strength observed in the
transverse response.  The amplitude to produce a high momentum nucleon
from OPE (middle and right panels in Fig.~\ref{fig:diagrams}) is 
$\propto - {\bm \sigma}_i \cdot {\bf k} \,  {\bm \sigma}_j \cdot {\bf k} \, {\bm \tau}_i \cdot {\bm \tau}_j \, v^\pi (k)/ {e}(k)$.
Alternatively, two low-momentum nucleons (the dominant part of the ground-state wave function)
can interact with the photon by exchanging a pion and produce two high-momentum nucleons by
sharing the momentum transfer between the nucleons in the struck pair (right panel 
in Fig.~\ref{fig:diagrams}).  These two processes are produced by the same physics 
(OPE in either the ground-state wave function or current), 
and can yield the same final state, hence they interfere.  
As shown in detail in Appendix~\ref{app:inter}, assuming the initial momenta 
${\bf p}$ and ${\bf p}^\prime$ are small compared to the momentum transfer,
%the sum of these two diagrams is 
%$ \propto {\bm \sigma}_i \cdot {\bf k} \, {\bm \sigma}_i \cdot {\bf q}\, 
%({\bm{ \sigma}}_j \cdot {\bf k})^2 \, ({\bm \tau}_i \cdot {\bm \tau}_j)^2 \sim  v_\pi^2 (k)$,
%the square of the one-pion-exchange potential, and is therefore positive. 
%This result is worked out in detail in Appendix ~\ref{app:inter}, where
one finds that the amplitude squared associated with the diagram in the middle panel is
$M_\pi^2\simeq [G_M^V (Q^2)]^2 \, \mu_V^2 \, q^2\, \widetilde{v}_\pi^{\,2}(k)$, where
$G_M^V (Q^2)$ and $\mu_V^2$ are the isovector magnetic form factor and magnetic
moment, respectively, and $\widetilde{v}_\pi(k)=- f_{ {\pi} NN}^2/( k^2 + m_\pi^2)/(3\, m_\pi^2)$.
The amplitude squared associated to the last diagrams instead reads 
$ M_{\rm int}^2 \  \simeq \,
	G_E^V (Q^2)\, G_M^V (Q^2) \,  \mu_V\,  q^2 \,\widetilde{v}_\pi^{\,2} (k)$.
	These findings imply that both the correlation and 
	interference terms are positive and add up contributively 
	to the leading term obtained under the one-body prescription.
% 
% Assuming the initial momenta 
% ${\bf p}$ and ${\bf p}^\prime$ are small compared to the momentum transfer, 
% \rocco{The following
% doesn't look right to me, but I have NOT looked at the spin-isospin algebra in appendix.
% The seagull term has something like ${\bm \sigma}_i \, {\bm \sigma}_j\cdot {\bf k}$ \dots
% it is not obvious why the term proportional
% to ${\bm \sigma}_i \cdot {\bf q} \, {\bm \sigma}_i\cdot {\bf k}={\bf k}\cdot {\bf q} +\cdots$
% is always positive \dots The other issue I am having is that this formula as is written
% would say that the enhancement vanishes at $q$=0, which is not the case \dots 
% the sum of these two diagrams is 
% $ \propto {\bm \sigma}_i \cdot {\bf k} \, {\bm \sigma}_i \cdot {\bf q}\, 
% ({\bm{ \sigma}}_j \cdot {\bf k})^2 \, ({\bm \tau}_i \cdot {\bm \tau}_j)^2 \sim  v_\pi^2 (k)$,
% the square of the one-pion-exchange potential, and is therefore positive.
% \joe{check and/or remove}
% }

The contributions of the second and third diagrams have the same sign and are 
comparable at moderate momenta in this simple picture. They also have a similar energy
dependence.  The spin-isospin algebra used to obtain this result is detailed in 
Appendix~\ref{app:inter}.  In later sections, we show
that this constructive interference between one- and two-body currents
in correlated nucleon pairs  persists
in the complete calculations based on the full ground state and
including full correlations along with the full set of one- and two-body currents.
We also examine the contribution in explicit final states with high momentum back-to-back pairs.

We conclude this section by observing that in the longitudinal response this
enhancement is largely absent, since the contributions associated with
two-nucleon charge operators are of relativistic origin, and quite
small at moderate momentum transfer (indeed, they vanish at vanishing
momentum transfer because of charge conservation).
In this case, though, the charge exchange process
from OPE (middle panel of Fig.~\ref{fig:diagrams}) 
leads to a smaller ``effective mass'' for the nucleon and thus spreads out the response,
reducing it near the peak~\cite{Carlson:1994zz,Carlson:2001mp}.

\section{QMC calculations of the response and scaling}
\label{sec:scaling}

Realistic interactions and currents, and the imaginary-time formalism have been used to 
calculate electroweak response functions, and associated electron~\cite{Carlson:2001mp,Lovato:2015qka,Lovato:2016gkq}
and neutrino~\cite{Lovato:2017cux} scattering. 
In such an approach, one evaluates the Laplace transform of the response,
\begin{equation}
\widetilde{R}_\alpha (q, \tau) = \int_{\omega_{\rm el}}^\infty d\omega\, {\rm e}^{- \omega  \tau }\, R_\alpha (q, \omega) \ ,
\end{equation}
which results in an imaginary-time response of the type
\begin{equation}
	\widetilde{R}_\alpha (q,\tau)= \overline{\sum_{M_i}}
   	     \langle \Psi_i | 
	  O_\alpha^\dagger ({\bf q}) \,{\rm e}^{-\left( H-E_i\right) \tau}\
	  O_\alpha ({\bf q}) | \Psi_i \rangle\ .
\label{eq:imaginarytime}
\end{equation}
Green's function Monte Carlo (GFMC) methods can then be used to calculate the relevant matrix elements~\cite{Carlson:2014vla}. 
Since the nuclear response in the quasielastic region is fairly smooth as a function of $\omega$,
maximum entropy techniques are successful in obtaining the real-time response from the
imaginary-time one~\cite{Lovato:2015qka}.  

This imaginary-time method has a major advantage over other currently available approaches.
Given a set of (local or minimally non-local) realistic interactions and accompanying electroweak
currents, it allows one to calculate {\it ab initio}, without any additional approximations beyond those
inherent to the modeling of these interactions and currents, the quasielastic response of a nucleus, by
treating completely correlations in the initial state, by accounting consistently through
the imaginary-time propagation for interaction effects in the final states, and, in particular, by retaining in full
the interference between one- and two-nucleon currents discussed in the previous section.
%(ifinal-state interactions  and two-nucleon currents are included completely and exactly; (ii) the two-nucleon interactions
%and corresponding currents are consistently constructed (see Ref.~\cite{Shen:2012}, and the review paper~\cite{Carlson:1998} for a more extended treatment); (iii) these interactions and currents
%(ii)  interactions and currents are tied to the same interaction used to calculate the ground state $|\Psi_0 \rangle$,
%and that the current operators are the same as those used to study other 
In this context, it is worthwhile pointing out that the realistic interactions and currents adopted in the
present work---Argonne $v_{18}$ two-nucleon~\cite{Wiringa:1994wb} and Urbana-IX three-nucleon~\cite{Pudliner95} interactions and associated currents~\cite{Shen:2012xz}---provide a quantitatively successful description of many nuclear electroweak
observables~\cite{Bacca:2014tla}, such as nuclear electromagnetic form factors~\cite{Schiavilla:2018udt,NevoDinur:2018hdo} 
and low-energy transitions including beta decays~\cite{Pastore:2012rp,Datar:2013pbd,Pastore:2014oda,Pastore:2017uwc}.  
They have also been used in
studies of double beta decay matrix elements~\cite{Pastore:2017ofx,Cirigliano:2018hja}.
The main disadvantages of this approach are that 
it is computationally intensive, since it propagates the full $A$-nucleon
system, and that it provides direct information only on the inclusive response, summed
over all final states.  Furthermore, its implementation is, at least for the time being,
limited to systems with mass number $A\le 12$.

Plane-wave impulse approximation (PWIA) approaches, either based on the momentum distribution
or spectral function, involve in essence off-diagonal density matrix elements of single nucleons.  Obviously,
they cannot preserve the full (non-energy-weighted) sum rules $S_\alpha(q)$, since two-nucleon currents
are ignored. They also underestimate energy-weighted sum rules associated with single-nucleon
currents~\cite{Schiavilla:1989jm}, which implies that the corresponding PWIA response functions
(obtained with single-nucleon currents) will not have the correct $\omega$ dependence.  More recently,
approaches have appeared based on factorization of the final $A$-nucleon state into a two-nucleon
plane-wave state and an $A$--2 spectral function~\cite{Rocco:2015cil}, which include both one- and two-nucleon currents.  

The GFMC calculations provide an explanation for the observed scaling of the electromagnetic
response, both the scaling with momentum transfer $q$ ($y$-scaling) and the
scaling with the mass of the target.  Since they reproduce the experimental results,
they obviously scale with momentum transfer.  We have argued in the previous section that in
the transverse channel the interference between one- and two-nucleon currents leads to
final states very similar to those of high-momentum nucleons induced by the interaction. This is observed
explicitly in the GFMC calculations, since the excess strength in the response is spread
out across the peak region
in a way similar to how the momentum distribution broadens the response obtained with single-nucleon
currents.  Hence $y$-scaling is preserved, even though it is not a purely one-body mechanism that
produces the cross section.  Note that the scaling in the longitudinal channel is quite
different than in the transverse one, since it does not include any significant interference.

The GFMC calculations also proceed by evaluating path integrals that, at high energies $E$,
correspond to short imaginary times.  The full sum rule is obtained from the imaginary
time response at $\tau\,$=$\,0$.  The free single-particle propagator is
a Gaussian $\propto \ \exp [ - ({\bf r}_i - {\bf r}'_i)^2  / (4 \tau \hbar^2/ 2 m)] $. At $\tau\,$=$\, 0.01\, {\rm MeV}^{-1}$ the nucleons have only
propagated a distance of approximately 
$ \simeq 1.1$ fm , and
for $\tau \,$=$\, 0.02$ MeV$^{-1}$ the equivalent distance $d$ is about $1.6$ fm.  Thus,
the propagation at short times entirely determines the high energy response at $E \gtrsim50$
MeV.  For these short imaginary times the operator describing the propagation is nearly local.
Over such short distances all atomic nuclei with $N \sim Z$ look alike in the interior
region~\cite{Donnelly:1999sw}.  Hence, the superscaling
describing the scaling with mass number is quite accurate.  Allowing slightly different surface
regions for small and large nuclei (allowing different effective Fermi momenta) improves
this agreement further.  These arguments do not depend upon the structure of the current
operator, only that the energies are near or above the Fermi energy.  Hence they should apply
equally well to neutrino and electron scattering.  

In the picture outlined above, scaling is violated in the resonance region because the simple
relationship between momentum (or distance) with energy is lost.  Slow-moving pions and
$\Delta$-resonances can be created, which will require a lot of energy but 
not large momentum.
Hence, simple scaling in this region is not expected nor is it observed experimentally.

\section{Real-Time Response at Short Times}
\label{sec:realtime}

The importance of the short imaginary-time propagation naturally leads to an approach
that incorporates as much of the quasielastic physics as possible by evaluating
path integrals of one- and two-nucleon currents in real time.  Such an approach, which
we will refer to as the short-time approximation (STA),
keeps the full sum rules, the physics of the momentum distribution and PWIA, and
the behavior of the two-nucleon correlations and currents at short times or high
energies, corresponding to the Fermi energy and above. While keeping consistently
two-body physics and ensuing quantum interference contributions, 
the STA will not yield the correct physics for 
low-lying excitations or collective behavior like giant resonances.

In the STA we evaluate the real-time matrix element in Eq.~(\ref{eq:realtime}) 
for short times by retaining the full ground state and current operators,
and final state interactions at the two-nucleon level---specifically, those final state
interactions affecting only pairs involved at the electromagnetic interaction vertex.  This short-time
approximation should be valid at high energy transfer $\omega$ and moderate-to-high
momentum transfer $q$.  It naturally incorporates two-nucleon interactions and currents
and their interference, all of which have been demonstrated to be important in the discussion
above and in many papers previously~\cite{Carlson:1994zz,Carlson:1997,Carlson:2001mp}.  Since it is based
on the full $A$-nucleon ground state, it also accounts for the statistical correlations 
implied by the Pauli principle (Pauli blocking).  The (non-energy-weighted) sum rules are recovered at $t\,$=$\,0$ in
the short-time approximation.  However, before illustrating the STA approach more in detail,
it is useful to discuss how the PWIA response follows from Eq.~(\ref{eq:realtime}).
% 
% 
% reproduces the correct nuclear non-energy-weighted 
% sum rules  for all momentum transfers.
% 
% At low momentum transfers, Eq.~(\ref{eq:realtime}) includes Pauli
% blocking as it is evaluated in the full
% $A$-nucleon ground state.  Furthermore, since the full currents and exact ground state are included
% in Eq. \ref{eq:realtime}, .
\subsection{Plane-Wave Impulse Approximation (PWIA) Response} 
In PWIA the many-nucleon propagator is simply approximated as
%\begin{equation}
%\label{eq:e12}
%\langle {\bf R}^\prime\, \alpha^\prime | {\rm e} ^{- i H t } | {\bf R}\,\alpha \rangle
%\approx
%\langle {\bf r}_i^\prime| {\rm e}^{-iH^0_i t} |{\bf r}_i\rangle \, \delta_{\alpha_i^\prime,\alpha_i}\ 
%	\langle {\bf R}_{A-1}^\prime\, \alpha^\prime_{A-1} | {\rm e} ^{- i E_{A-1} t } | {\bf R}_{A-1} \,\alpha_{A-1} \rangle\ ,
%\end{equation}
\begin{equation}
\label{eq:e12}
\langle {\bf R}^\prime\, \alpha^\prime | {\rm e} ^{- i H t } | {\bf R}\,\alpha \rangle
\approx
\langle {\bf r}_i^\prime| {\rm e}^{-i(H^0_i+E_{A-1})t} |{\bf r}_i\rangle \, \delta_{\alpha_i^\prime,\alpha_i}\,
\prod_{k\ne i} \delta({\bf r}^\prime_k-{\bf r}_k)\, \delta_{\alpha^\prime_k,\alpha_k} \ ,
%	\langle {\bf R}_{A-1}^\prime\, \alpha^\prime_{A-1} | {\rm e} ^{- i E_{A-1} t } | {\bf R}_{A-1} \,\alpha_{A-1} \rangle\ ,
\end{equation}
where nucleon $i$ with kinetic energy $H^0_i$ is the struck nucleon, and the remaining $A-1$ nucleons
are treated as static spectators.  The $A$ nucleon spatial,
spin and isospin states are collectively denoted as 
${\bf R}^\prime\, \alpha^\prime$ and ${\bf R}\, \alpha$, where ${\bf R}\, \alpha\,$=$\,
({\bf r}_1 \, \alpha_1, \dots, {\bf r}_A\, \alpha_A)$ and similarly for the primed variables
($\alpha_i=\sigma_i\,\tau_i$ specify the spin and isospin states of nucleon $i$).
The constant $E_{A-1}$ shifts the energy of the response, and
can be interpreted as an average removal energy.
%while the spatial matrix elements fixes the coordinates of the $A-1$ nucleons.
Thus, the PWIA is related to the off-diagonal one-body density matrix.
 In the naive limit of Eq.~(\ref{eq:e12}), the current-current correlator, 
schematically denoted as
$\langle O^\dagger\, O\rangle$,  is given by
\begin{eqnarray}
&& \langle O^\dagger\, O\rangle \Big|_{\rm PWIA}
=\sum_i \sum_{\alpha_i \alpha_{A-1}} \int d{\bf r}_i^\prime\,d{\bf r}_i d{\bf R}_{A-1} \,
 \langle \Psi_i| O_i^\dagger({\bf q})|
 {\bf r}_i^\prime\,\alpha_i,  {\bf R}_{A-1}\,\alpha_{A-1} \rangle\,
 \langle {\bf r}_i^\prime| {\rm e}^{-iH^0_i t} |{\bf r}_i\rangle \nonumber\\
 &&\hspace{6cm}  \langle
 {\bf r}_i\,\alpha_i,  {\bf R}_{A-1}\,\alpha_{A-1} | O_i({\bf q})|\Psi_i \rangle\ ,
 \label{eq:eb7}
 \end{eqnarray}
where ${\bf R}_{A-1}\, \alpha_{A-1}$ is a short-hand notation for the spatial
and spin-isospin states of the spectator nucleons.
%\begin{equation}
%\langle O^\dagger\, O\rangle \Big|_{\rm PWIA} = 
%\sum_i 
% \sum_{\alpha} \int d{\bf R}^\prime\, d{\bf R}\,  
%	\langle \Psi |O^\dagger_i ({\bf q})| {\bf R}^\prime\,\alpha^\prime\rangle%
%	\langle {\bf r}_i^\prime | {\rm e}^{-i(H^0_i+E_{A-1})  t} |{\bf r}_i\rangle  
%	\langle {\bf R}\,\alpha |O_i({\bf q}) | \Psi\rangle 
%	\label{eq:eb7}
%\end{equation}
Only the one-body terms in which the same nucleon $i$ is involved (the ``active
nucleon'') are kept, all remaining terms in $O({\bf q})$ are ignored.
The eigenstates of the one-particle system are simple plane waves, and the
expression above yields a response depending only upon the single-particle
momentum distribution.
A better
treatment would require keeping the propagating eigenstates of the $A-1$ system,
which would lead to a similar
expression as in Eq.~(\ref{eq:eb7}), except for the spectral function replacing the
momentum distribution and for the presence of an additional integration over the (removal) energy.

We conclude this brief review of the PWIA by noting that the corresponding sum rule is obtained as
\begin{equation}
S^{\rm PWIA}(q) = {\rm Tr}\left[O_i^\dagger({\bf q})\, O_i({\bf q}) \right]/G^2(Q^2_{\rm qe}) \ ,
\end{equation}
where the trace is over the spin-isospin states of a single nucleon.
For example, in the longitudinal channel, $O_i^{L}({\bf q})$ is
given by (up to relativistic corrections proportional to $1/m^2$) 
\begin{equation}
O_i^{L}({\bf q})=\left[ G_E^p(Q^2_{\rm qe})\, P_{i,p }+G_E^n(Q^2_{\rm qe})\, P_{i,n} \right]\, {\rm e}^{i{\bf q}\cdot{\bf r}_i}\ ,
\end{equation}
where the $G_E^p$ and $G_E^n$ are the proton and neutron electric form factors, and $P_{i,p/n}$ is
the proton/neutron projector.  Thus, we find $S_L^{\rm PWIA}(q) \,$=$\,1$,
where we have taken $G_L^2=G_E^{p\, 2}+G_E^{n\,2}$.
Pauli blocking terms, particularly in medium- to heavy-weight nuclei, will reduce this
sum rule at low $q$, while (in the transverse channel) two-nucleon physics will increase
it at larger $q$.
\subsection{Short-Time Approximation (STA) response}

The STA includes the two-nucleon contributions that are ignored in the PWIA and, in
particular, accounts for the interference between one- and two-nucleon 
currents.  It is explicitly constructed as a function of both the momentum
transfer $q$ and energy transfer $\omega$, and hence must be calculated
separately for different $q$ as a full two-nucleon off-diagonal matrix
in the $A$-body system.  In the STA, the current-current correlator is approximated as
\begin{eqnarray}
O^\dagger\, {\rm e}^{-iHt}\,  O
	&= & \left( \sum_i O_i^\dagger + \sum_{i<j} O_{ij}^\dagger \right)
	{\rm e}^{-i H t} 
	\left(	\sum_{i'} O_{i^\prime} + \sum_{i'<j'} O_{i'j'} \right)
	 \nonumber \\
&=&\sum_i O_i^\dagger\,  {\rm e}^{-iHt} O_i 
	+ \sum_{i \neq j} O_i^\dagger\,  {\rm e}^{-iHt} O_j \nonumber\\
	&&+ \sum_{i\neq j}
	\left( O^\dagger_{i}\,{\rm e}^{-iHt}\, O_{ij} + O^\dagger_{ij}\,{\rm e}^{-iHt}\, O_{i}   + O^\dagger_{ij} \,{\rm e}^{-iHt}\, O_{ij}\right)+\cdots \ ,
\label{eq:e25}
\end{eqnarray}
dropping terms with three-or more active nucleons (the $\cdots$ in the above equation).  In particular, the Hamiltonian
$H$ only includes two-nucleon interactions. Three-nucleon
interaction effects are ignored in the propagation and therefore in the final states,
although they are included in the ground state.  In nuclear ground states, expectation values of
three-nucleon interactions are typically 5--10\% than those of two-nucleon interactions.

In the STA we assume that only the active pair (say, pair $ij$) propagates, and therefore the $A$-nucleon
propagator is approximated as
\begin{equation}
 \langle {\bf R}^\prime\, \alpha^\prime| {\rm e}^{-iH t} |{\bf R}\, \alpha\rangle\approx
\langle {\bf R}^\prime_{ij} |{\rm e}^{-i H^{\rm cm}_{ij} t} |{\bf R}_{ij}\rangle \,
\langle {\bf r}^\prime_{ij} \,\alpha_i^\prime\, \alpha_j^\prime 
| {\rm e}^{-iH^{\rm rel}_{ij} t} 
|{\bf r}_{ij} \,\alpha_i\,\alpha_j\rangle 
	\prod_{k \neq (i,j) }^A \delta({\bf r}_k^\prime-{\bf r}_k)\, \delta_{\alpha_k^\prime,\alpha_k}\ ,
\end{equation}
where $H^{\rm cm}_{ij}={\bf P}^2_{ij}/(4m)$ and $H^{\rm rel}_{ij}={\bf p}^2_{ij}/m+v_{ij}$
are, respectively, the center-of-mass and relative Hamiltonians (hereafter, unless
necessary for clarity, the active pair subscripts will be understood, for example
${\bf R}^\prime_{ij}\rightarrow {\bf R}^\prime$ and so on).  For the purpose of illustration,
in the following we only discuss
in detail the terms in Eq.~(\ref{eq:e25}) that lead to interference; we treat the
incoherent terms---first sum in this equation---similarly, but will not discuss them any further below.  We proceed as in the previous
section by introducing a complete set of position and spin-isospin states for the $A$-nucleon system,
which allows us to express the coherent terms  as (the active pair is $ij\,$=$\,12$)
\begin{eqnarray}
&& \langle O_L^\dagger\, O_R\rangle \Big|_{\rm STA}
= \frac{A(A-1)}{2}\sum_{\alpha_1^\prime \alpha_2^\prime \alpha_1 \alpha_2}\,\, \sum_{\alpha_{A-2}}
\int d{\bf R}^\prime d{\bf r}^\prime  \,d{\bf R} \, d{\bf r} \,d{\bf R}_{A-2}\nonumber\\
&& \times\langle \Psi_i| O_L^\dagger({\bf q})|
 {\bf R}^\prime,{\bf r}^\prime\, \alpha^\prime_1\,\alpha^\prime_2,  {\bf R}_{A-2}\,\alpha_{A-2}\rangle\,
\langle {\bf R}^\prime |{\rm e}^{-i H^{\rm cm}_{12} t} |{\bf R}\rangle \,\nonumber\\
&&
\times\langle {\bf r}^\prime\,\alpha_1^\prime\, \alpha_2^\prime 
| {\rm e}^{-iH^{\rm rel}_{12} t} 
|{\bf r} \,\alpha_1\,\alpha_2\rangle\,
 \langle {\bf R},{\bf r}\, \alpha_1\,\alpha_2,  {\bf R}_{A-2}\,\alpha_{A-2} |
 O_R({\bf q})|
 \Psi_i\rangle \ .
\label{eq:incoherent}
 \end{eqnarray}
The possible combinations for coherent contributions in $O_L^\dagger O_R$ (operators acting on 
the left and right wave functions) are $ \left( O_i^\dagger O_j\, ,  O_i^\dagger O_{ij}\, ,
O_j^\dagger O_{ij}  \right)$ and their adjoints.

We are then left with
the evaluation of the two-nucleon propagator, for which we use the following expression
obtained by summing over the bound and continuum eigenstates of $H_{12}^{\rm rel}$
 \begin{eqnarray}
\label{eq:ec7prime}
 \langle {\bf r}^\prime\, \alpha_1^\prime\, \alpha_2^\prime |   {\rm e}^{-iH^{\rm rel}_{12} t}|{\bf r}
 \, \alpha_1\, \alpha_2 \rangle &=&\sum_{\gamma} \int_0^\infty \!\! d e \,{\rm e}^{-i\,e\, t}\,\,
 \phi^{\gamma}_{\alpha_1^\prime\, \alpha_2^\prime}({\bf r}^\prime; e)\,
 \phi_{\alpha_1\, \alpha_2}^{\gamma\,*}({\bf r};e) \nonumber\\
&&+\, {\rm e}^{-i \,e_d\, t} \sum_{M_d=0,\pm1} \phi^{\gamma_d,M_d}_{\alpha_1^\prime\, \alpha_2^\prime} ({\bf r}^\prime;e_d)\,
 \phi^{\gamma_d,M_d\,*}_{\alpha_1\, \alpha_2} ({\bf r};e_d) \ ,
\end{eqnarray}
where $\gamma$ denotes the discrete quantum numbers that specify the continuum state,
namely $\gamma=JM_J,TM_T,SLL^\prime$ where $JM_J$ are the total angular
momentum and its projection along the quantization axis, $TM_T$ are the pair isospin and
isospin projection, and $SLL^\prime$ are, respectively, the pair spin, and incoming and outgoing
orbital angular momenta, while $\gamma_d$ specifies the quantum numbers of the
bound state (the deuteron), which occurs in channel $J,TM_T,S=1,00,1$ with
$e_d=-2.225$ MeV, and $M_d$ are the projections of
the total angular momentum.  In a less compact notation,
the continuum state, as an example, reads
\begin{equation}
\phi^{\gamma}({\bf r}; e)=
\frac{w^{JST}_{L^\prime L}(r;e)}{r} \, Y_{L^\prime SJ}^{M_J}(\hat{\bf r})\, \eta^T_{M_T} \ ,
\end{equation}
where $w^{JST}_{L^\prime L}(r;e)$ are solutions of the radial Schr\"odinger equation in channel
$JST$ with relative energy $e$, $Y_{LSJ}^{M_J}$ are standard spin-angle functions, and $\eta^T_{M_T}$
are isospin states with $T M_T$, and $\phi^\gamma_{\alpha_1\alpha_2}$ denotes
the projection of $\phi^\gamma$ on the individual spin-isospin states $\alpha_1\alpha_2$
of the active pair.  In the present calculations, interaction effects (in the active pair) are
included exactly for all $J$ with $J \le J_{\max}\,$=$10$. For $J>J_{\max}$ the continuum solutions
are replaced by spherical Bessel functions, that is 
\begin{equation}
\label{eq:e34}
\frac{w^{JST}_{L^\prime L}(r;e)}{r} \longrightarrow \delta_{LL^\prime}\, j_L(\sqrt{me}\, r) \qquad {\rm for}\,\, J > J_{\max} \ .
\end{equation}

It is convenient to express the STA response,
which now includes also the contribution of the incoherent term in Eq.~(\ref{eq:e25}),
as an integral over the center-of-mass and relative energies,
\begin{equation}
 \label{eq:density.formula.e}
R^{\rm STA}(q,\omega) = \int_0^\infty \, de  \int_0^\infty  dE_{\rm cm} \, \,
\delta \left(\omega+E_i-e - E_{\rm cm} \right) \,   D(e, E_{\rm cm})  \ ,
\end{equation}
and the function $D(e,E_{\rm cm})$ can be obtained from Eq.~(\ref{eq:incoherent}) (the resultin
expression of course
includes the factors arising from the change of variables $P \rightarrow E_{\rm cm}$
and the integration over the solid angle specified by the ${\bf P}$-direction).  It is
worthwhile pointing out here that one could easily account for the dependence on the direction of the relative
momentum ${\bf p}$ (rather than just its magnitude) by expanding the
two-nucleon propagator in Eq.~(\ref{eq:ec7prime}) in terms of continuum states specified
by the relative momentum ${\bf p}$ and pair spin-isospin states $SM_STM_T$ (see Ref.~\cite{Shen:2012xz}).

The factorization outlined above retains fully interaction effects at the two-nucleon level, and
accounts for the crucial interference between one- and two-body terms in the electromagnetic
current operator.  Small contributions involving three or more active nucleons as well as
interactions between the active pair and the remainder of the nucleus are neglected.
As a consequence, the present approach will not produce the correct threshold
behavior for the response,  but will
reflect that of the underlying two-body physics.  For example, at low momentum transfer
($q \lesssim 300$ MeV/c) the STA transverse response for a nucleus will contain a peak
in the threshold region associated with the magnetic transition from the quasi-deuteron
state---a pair of nucleons in spin-isospin $ST\,$=$\,10$ in the ground state of the nucleus---to
the quasi-bound state---a pair of nucleons in relative S-wave and spin-isospin $ST\,$=$\,01$.
This peak is seen in calculations of the transverse response of the deuteron~\cite{Shen:2012xz},
but it is an artifact here.  Simple estimates can be parametrized to take care
of this issue (see below).  Lastly, up to factors of $G^2_\alpha$ in Eq.~(\ref{eq:sumrule}),
the (non-energy weighted) sum rule results from integrating the $D$-function over the relative
and center-of-mass energies.

\subsection{STA response densities as function of center-of-mass and relative energies}

\begin{figure}[!htbp]
\centering
\begin{minipage}{.5\textwidth}
\centering
\includegraphics[height=2.2in]{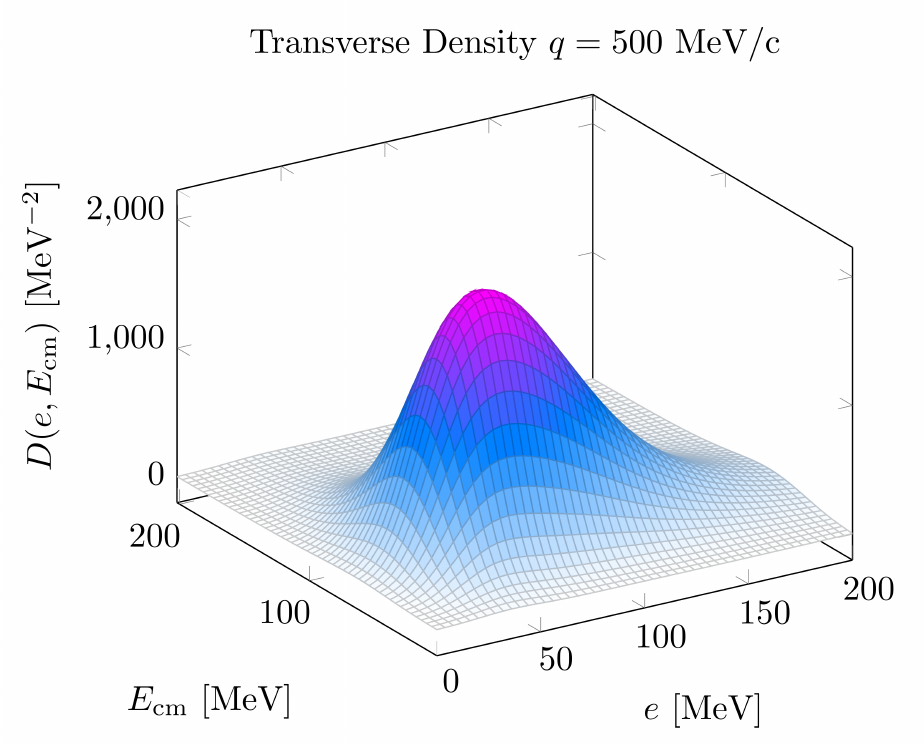}
\end{minipage}%
\begin{minipage}{0.5\textwidth}
\centering
\includegraphics[height=2.2in]{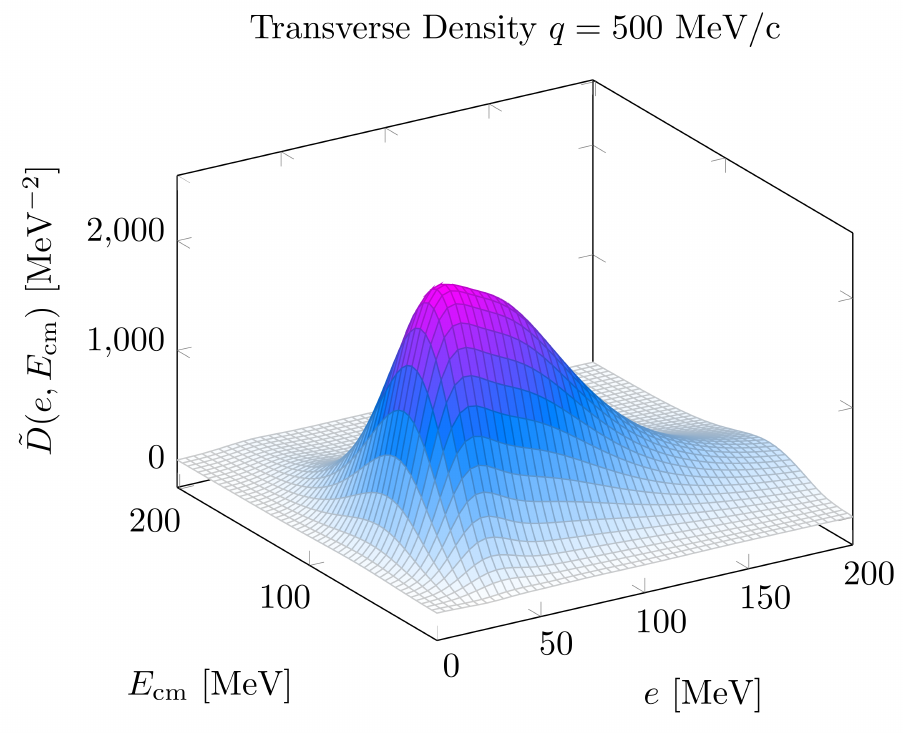}
\end{minipage}
\caption{\small Alpha particle transverse response densities at  $q\,$=$\,500$ MeV/c.  
The surface plots
show the response densities as functions of relative energy $e$ and center-of-mass energy $E_{\rm cm}$.
Results prior (left panel) and after (right panel) the shift (see text).
}
\label{fig:3dsurf}
\end{figure}

The response densities can be obtained as
a function of the pair relative energy $e$ and center-of-mass energy $E_{\rm cm}$
after the interaction vertex with the virtual photon (see Fig. \ref{fig:3dsurf}). 
As noted in the previous section, in principle more detailed information on angles could also be kept. 
The expected long tail in relative energy, induced by two-body physics, is apparent in the figure. 
The response function at a given momentum and energy transfer $q$ and $\omega$
is given as an integral over the response density  $D(e,E_{\rm cm})$ at a given
$q$ for $\omega +E_i= e+ E_{\rm cm}$. In the next section we compare results
for the response functions obtained in the 
short-time approximation to the imaginary-time calculations.

The STA does not have any knowledge of the correct threshold 
behavior or low-energy properties of the system.  Since the sum rules are obtained accuratly,
we can include our knowledge of the thresholds by making the replacement
\begin{equation}
	\widetilde{D}( e^\prime, E_{\rm cm} ) = \int_0^\infty de \, D(e, E_{\rm cm} ) \,
	N  \, {\rm exp}\!  \left\{ - \left[\frac{ e^\prime- \omega(e)} {\omega_{\rm th}}\right]^2
	\right\}\ ,
\end{equation}
with
\begin{equation}
\omega (e) = \sqrt { e^2 + {\omega_{\rm th}}^2 \ \exp( -e /\overline{\omega}) } \ ,\qquad
	\int_0^\infty de\,  N(e) \, {\rm exp}\!\left\{ - \left[\frac{ e^\prime- \omega(e)} {\omega_{\rm th}}\right]^2\right\}= 1\ ,
\end{equation}
the last condition defines $N$ to ensure the (non-energy) sum rules are preserved.  This transformation shifts the very low energy response by
$\omega_{\rm th}$ and distributes it with a Gaussian distribution.  The two parameters controlling
its shift and width are $\omega_{\rm th}$ and $\overline{\omega}$.  For $ ^{4}$He we use $\omega_{\rm th}\,$=$\, 35$ MeV
and a width $\overline{\omega} \,$=$\, 15$ MeV. With these choices there is very little strength below
the physical threshold of $\sim 20$ MeV.  The response densities before and after the shift are
illustrated in Fig. \ref{fig:3dsurf}.  One could also add an extra mean-field potential to the
two-nucleon system to mimic the average impact of interactions with the spectator nucleons.
Finally, the energy-weighted sum rule $W^{\rm STA}(q)$ is obtained (after the shift) as
\begin{eqnarray}
G^2(Q^2_{\rm el})\, W^{\rm STA}(q) &=&\int^\infty_{\omega_{\rm el}} d\omega \,\omega\, R^{\rm STA}(q,\omega) \nonumber\\
&=&\int_0^\infty \, de^\prime  \int_0^\infty  dE_{\rm cm} \,  
\left(e^\prime + E_{\rm cm} \right) \,   \widetilde{D}(e^\prime, E_{\rm cm}) \ . 
\end{eqnarray}
The two-nucleon final states include corrections of 
order $v_{ij} \, t$ to the propagator, which make important contributions to the
energy-weighted sum rule.  

\section{Results for Inclusive Scattering}
\label{sec:inclusiveresults}

In this section, we summarize the response calculations for inclusive electron scattering
on $^4$He, comparing our results to full GFMC results~\cite{Carlson:2001mp,Lovato:2015qka} 
and experimental data.   A description of the two-body charge and current operators used in this work
is provided most recently in Refs.~\cite{Shen:2012xz,Lovato:2015qka} and references therein. 
The STA should work well for momentum transfers greater than the Fermi momentum
and for energy transfers above the giant resonance and below the excitation energy of $\Delta$ and higher-lying resonances.

First, we report the numerical values for the longitudinal and transverse sum rules
obtained by (i) integrating the STA response functions (S$_{L/T}^{\rm STA}$ in 
Table~\ref{tb:sumrule}), (ii) calculating the sum rules ``exactly'' within STA (S$_L^{\rm STA*}$) that
is ignoring terms involving three and four nucleons, and (iii) including all terms in the
sum rule calculation (S$_{L/T}$).
We compare our results to the full GFMC calculations of Ref.~\cite{Carlson:2001mp}.
The STA preserves the sum rules.  Since we are calculating the response densities
up to finite maximum center-of-mass and relative energies the agreement between
integrating over the response density and by direct evaluations of the sum rule (in the STA limit)
is not exact.  The one-body current sum rules are reproduced within
a few percent, but the full sum rules including two-nucleon currents are
somewhat smaller in the integrated response density due to the fact that the high relative
energy piece of the response is cut off.  As a matter of fact, the STA calculations 
are carried out up to relative and center-of-mass energies of $\sim \,$800 MeV. Increasing 
the range of the pair energies would improve the agreement with the exact 
estimates. 

\begin{table}
\begin{tabular}{|c||c|c|c|c||c|c|c|c|}
\hline
$q$ [MeV/c]& S$_L^{\rm STA}$     & S$_L^{\rm STA*}$ & $S_L$      & $S_L$ Ref.\cite{Carlson:2001mp}   & $S_T$ & S$_T^{\rm STA*}$ & $S_T$ & $S_T$ Ref.~\cite{Carlson:2001mp} \\
\hline
300        & (0.59)0.60          & (0.66)0.67       & 0.66(0.65) & (0.67)0.65    & (0.83)1.33 & (0.88)1.54  & 0.89(1.53)   &  (0.91)1.58  \\  
\hline                                                                                                       
400        & (0.80)0.79          & (0.83)0.82       & 0.83(0.81) & (0.86)0.81    & (0.93)1.34 & (0.95)1.47  & 0.97(1.48)   &  (0.98)1.50  \\   
\hline                                                                                                       
500        & (0.87)0.86          & (0.88)0.87       & 0.89(0.88) & (0.94)0.88    & (0.98)1.34 & (1.00)1.43  & 1.00(1.43)   &  (1.01)1.44  \\   
\hline                                                                                                       
600        & (0.88)0.87          & (0.88)0.89       & 0.91(0.90) & (0.97)0.91    & (1.02)1.32 & (1.03)1.40  & 1.01(1.38)   &  (1.01)1.38  \\   
\hline                                                                                                       
700        & (0.87)0.88          & (0.88)0.89       & 0.92(0.92) & (0.99)0.94    & (1.00)1.32 & (1.07)1.40  & 1.01(1.34)   &  (1.01)1.33  \\  
\hline                                                                                                       
800        & (0.86)0.88          & (0.87)0.89       & 0.92(0.94) & --            & (1.08)1.33 & (1.10)1.41  & 1.01(1.31)   &  --          \\  
\hline
\end{tabular}
\caption{$^4$He Longitudinal and transverse sum rules obtained
by integrating the STA response, denoted as S$_{L/T}^{\rm STA}$,
and by direct evaluation of the current-current matrix element in
Eq.~(\ref{eq:sumrule}) but ignoring, however, three- and four-nucleon terms,
	denoted as S$_{\rm L/T}^{\rm STA*}$. S$_{L/T}$ are the full sum rules.
	These results are
compared with those eported in Tables I and III of Ref.~\cite{Carlson:2001mp} and referred to as
S$_{L/T}$ (Ref.~\cite{Carlson:2001mp}. 
Values in parentheses are with one-body currents alone. The longitudinal sum rule is obtained
	by subtracting the elastic response.}
\label{tb:sumrule}
\end{table}

\begin{figure}[!htbp]
    \centering
    \begin{minipage}{.5\textwidth}
        \centering
        \includegraphics[height=2.in]{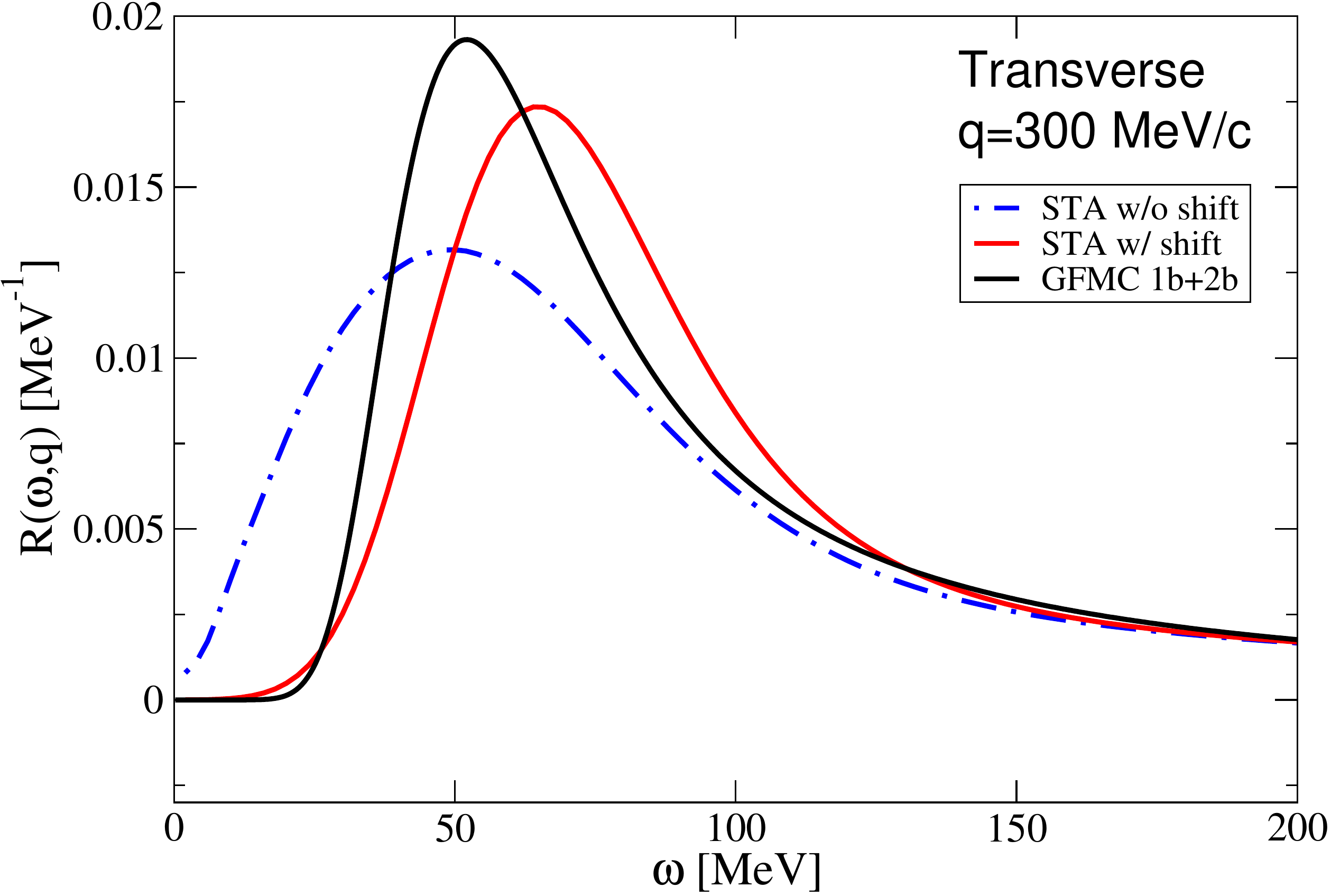}
    \end{minipage}%
    \begin{minipage}{0.5\textwidth}
        \centering
        \includegraphics[height=2.in]{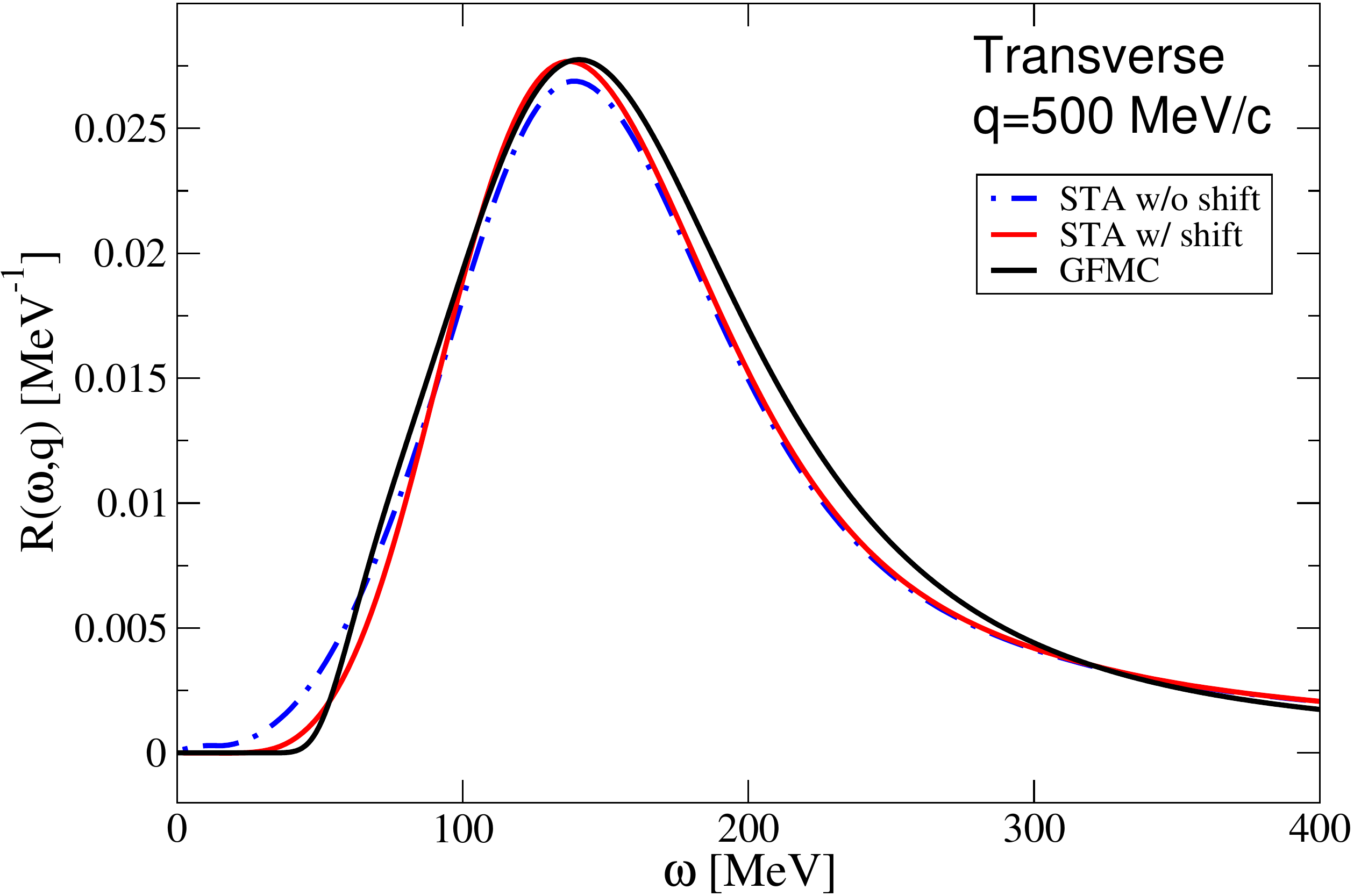}
    \end{minipage}
    \caption{\small STA results for $^4$He$(e,e^\prime)$ inclusive scattering. 
	Left panel: Transverse response at $q\,$=$\,300$ MeV/c, prior to shift (dashed-dotted  blue line) 
	and including shift (solid red line) compared with the exact GFMC response (solid black line).
	Right panel: Same but at $q\,$=$\,500$ MeV/c. See text for further explanations.
}
    \label{fig:compareGFMC}
\end{figure}

\begin{figure}[!htbp]
    \centering
%     \begin{minipage}{.5\textwidth}
%         \centering
%         \includegraphics[height=2.in]{STAvsGFMC-Long-300.pdf}
%     \end{minipage}%
%     \begin{minipage}{0.5\textwidth}
        \centering
        \includegraphics[height=2.5in]{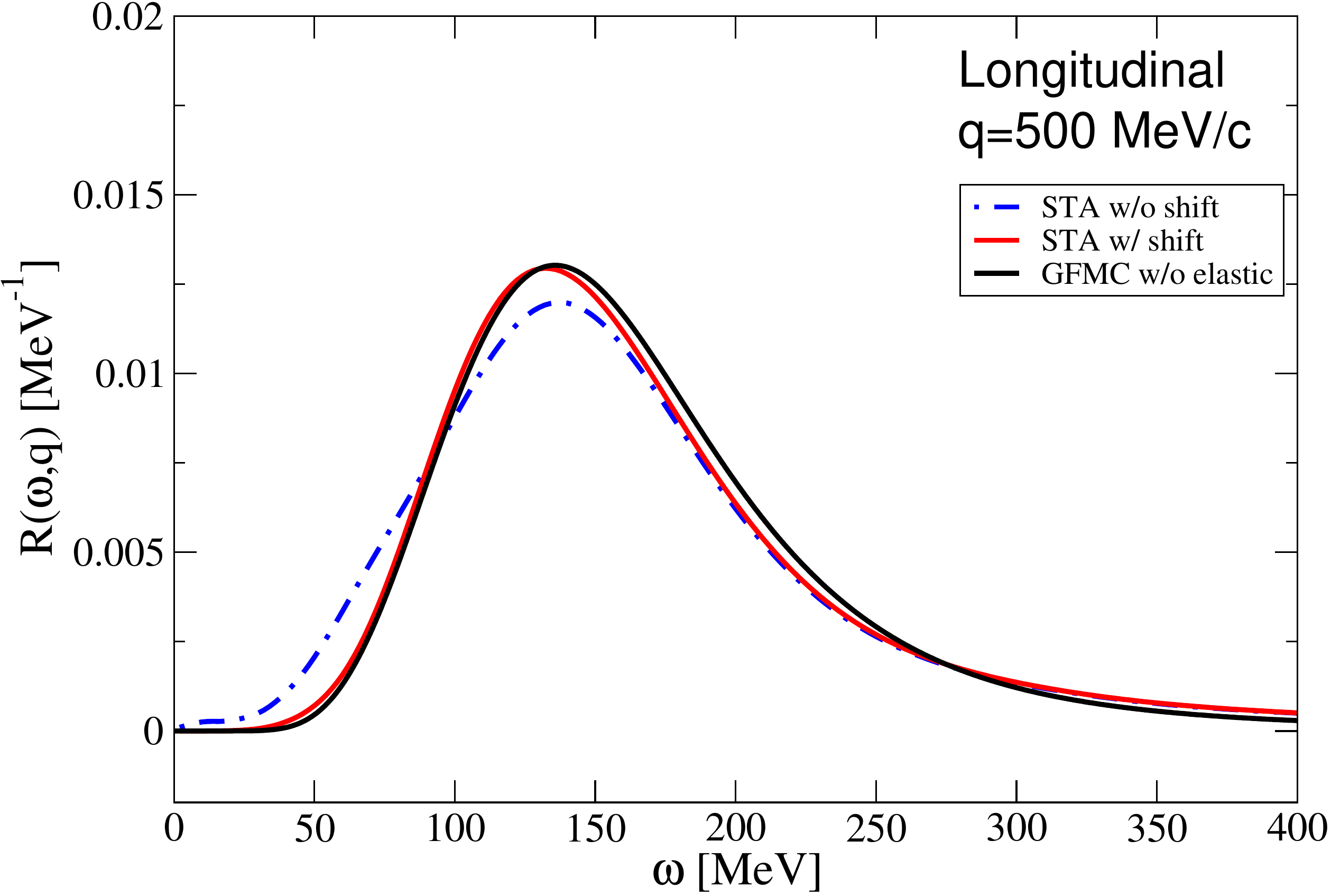}
%     \end{minipage}
    \caption{\small STA results for $^4$He$(e,e^\prime)$ inclusive scattering.  
	Longitudinal response at $q=500$ MeV/c. Notation as in Fig.~\ref{fig:compareGFMC}.
}
    \label{fig:compareGFMCL}
\end{figure}

The total transverse response compared to GFMC is shown in Fig.~\ref{fig:compareGFMC}.  
At $q = 300$ MeV/c a direct
evaluation of the STA as described above puts too
much strength at very low energies below the physical threshold.
The dashed line in Fig.~\ref{fig:compareGFMC} shows the response obtained without any knowledge
of the threshold while the full red line shows the results obtained by enforcing the correct
behavior at threshold, as discussed in previous section.  This response is
in pretty good agreement with the GFMC results.  It is unlikely the 
STA alone would be useful in heavier nuclei 
below $300$ MeV/c, as giant
resonances and other low-lying states start to dominate.  It could
perhaps be extended by combining the imaginary- and real-time response
approaches and by calculating the inverse
energy weighted sum rule (susceptibility).
The transverse and longitudinal STA responses are compared with corresponding
GFMC ones in Figs.~\ref{fig:compareGFMC} and~\ref{fig:compareGFMCL}. 
At this higher momentum transfer, the shift has little impact and the STA
response is a faithful reproduction of the (exact) GFMC response.
Both the GFMC and STA results for the longitudinal response of $^4$He
are also in good agreement  with the {\it ab initio} LIT calculations by Bacca
{\it et al.}~\cite{Bacca:2008tb}.

\begin{figure}[!htbp]
    \centering
    \begin{minipage}{.5\textwidth}
        \centering
        \includegraphics[height=2.2in]{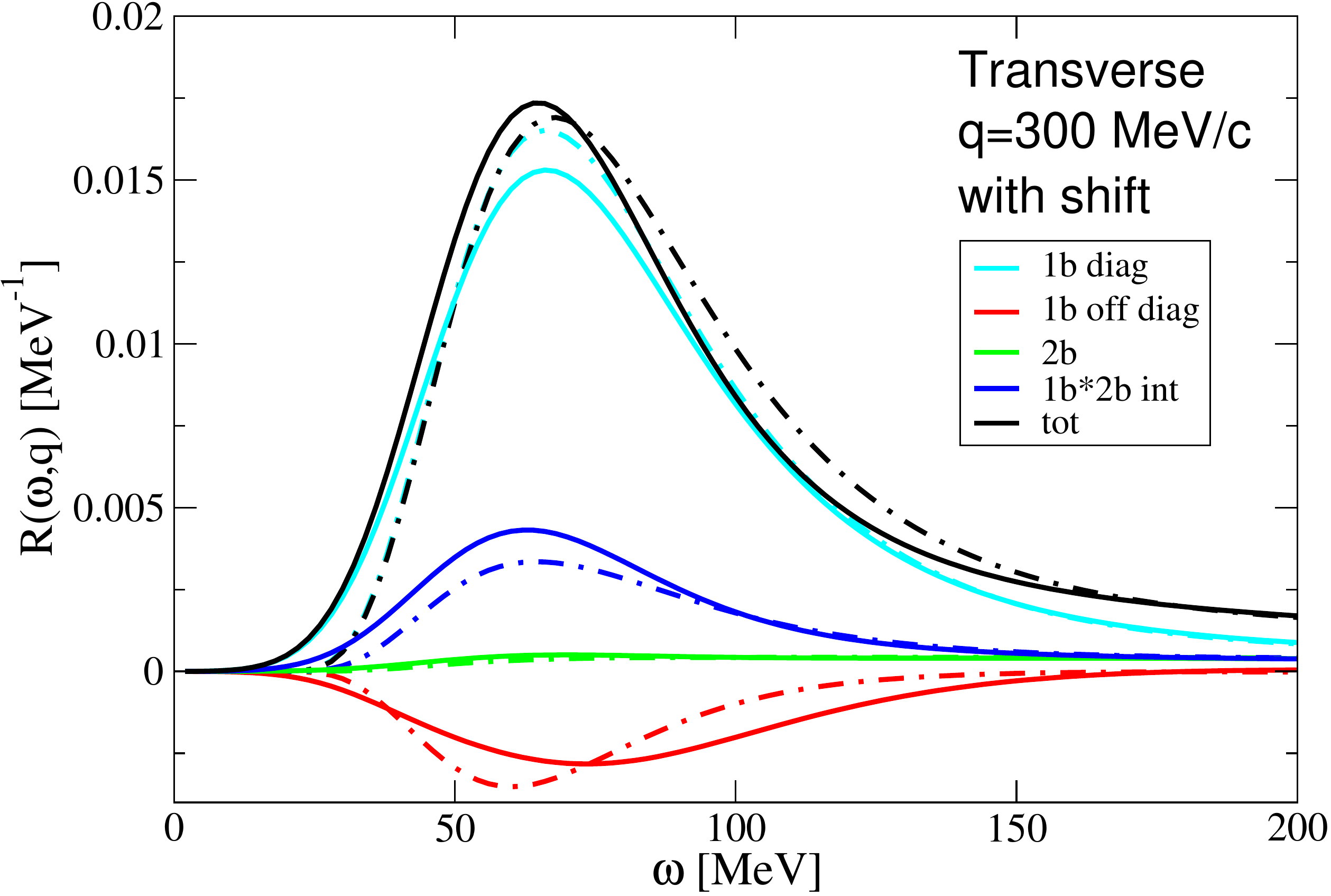}
    \end{minipage}%
    \begin{minipage}{0.5\textwidth}
        \centering
	    \includegraphics[height=2.2in]{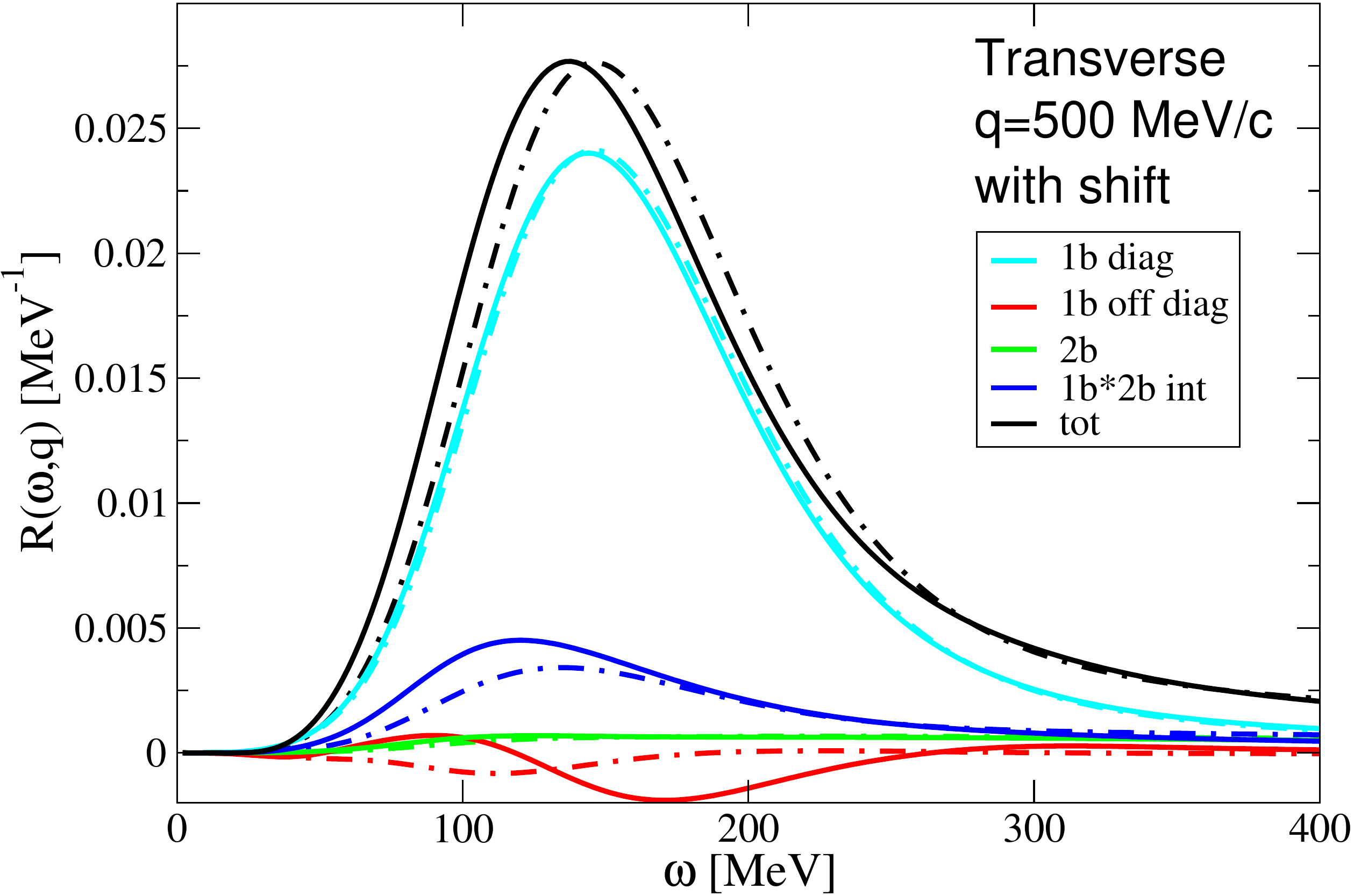}
    \end{minipage}
	\caption{ \small Comparison of  transverse responses without (dashed lines) and with (full lines) 
	interacting two-nucleon final states. Various contributions are shown, including one-body current
	diagonal terms, one-body current off-diagonal ($i \neq j$) terms, interference between one- and
	two-body currents, and two-body currents only.
	See text for further explanations.
	Results at $q\,$=$\,300$ MeV/c (left panel) and $q\,$=$\,500$ MeV/c (right panel).
}
	\label{fig:freevsfull}
\end{figure}

We can gauge the impact of the final-state interactions within the pair
by comparing results obtained with the interacting two-nucleon propagator to those
obtained with the free-particle propagator via the
replacement in Eq.~(\ref{eq:e34}).  As shown in Fig.~\ref{fig:freevsfull}, the final state
interactions within the pair 
at $q\,$=$\, 300$ and $500$ MeV/c shift strength to lower energies.  
At low energy, this is especially apparent before the inclusion of the 
shift in $\omega$ via the inclusion of the threshold $\omega_{\rm th}$.

The response can also be divided into one-body diagonal or incoherent terms 
(those where the same single-nucleon current operator acts at the initial and final times, namely
terms of the type
$ {O}^\dagger_i \cdots {O}_i$), off-diagonal one-body terms
(one-body current operators from different particles, that is, $ {O}^\dagger_i \cdots {O}_j$),  
terms from the interference of one- and two-body currents 
(obtained by keeping terms of the type ${ O}_{ij}^\dagger \cdots O_i$ and
$ {O}_i^\dagger \cdots {O}_{ij}$), and two-body diagonal terms (proportional to $ O_{ij}^\dagger\cdots O_{ij}$).   
The contribution of the various terms are also shown in  Fig.~\ref{fig:freevsfull} for the transverse
response at different kinematics. As stated above,  we are ignoring terms involving currents operating on three or four
different nucleon coordinates in the interference and two-body off-diagonal pieces of the
response. These terms have been demonstrated to be small in the imaginary-time response
calculations.  

\begin{figure}[bth]
\centering
\includegraphics[width=\linewidth]{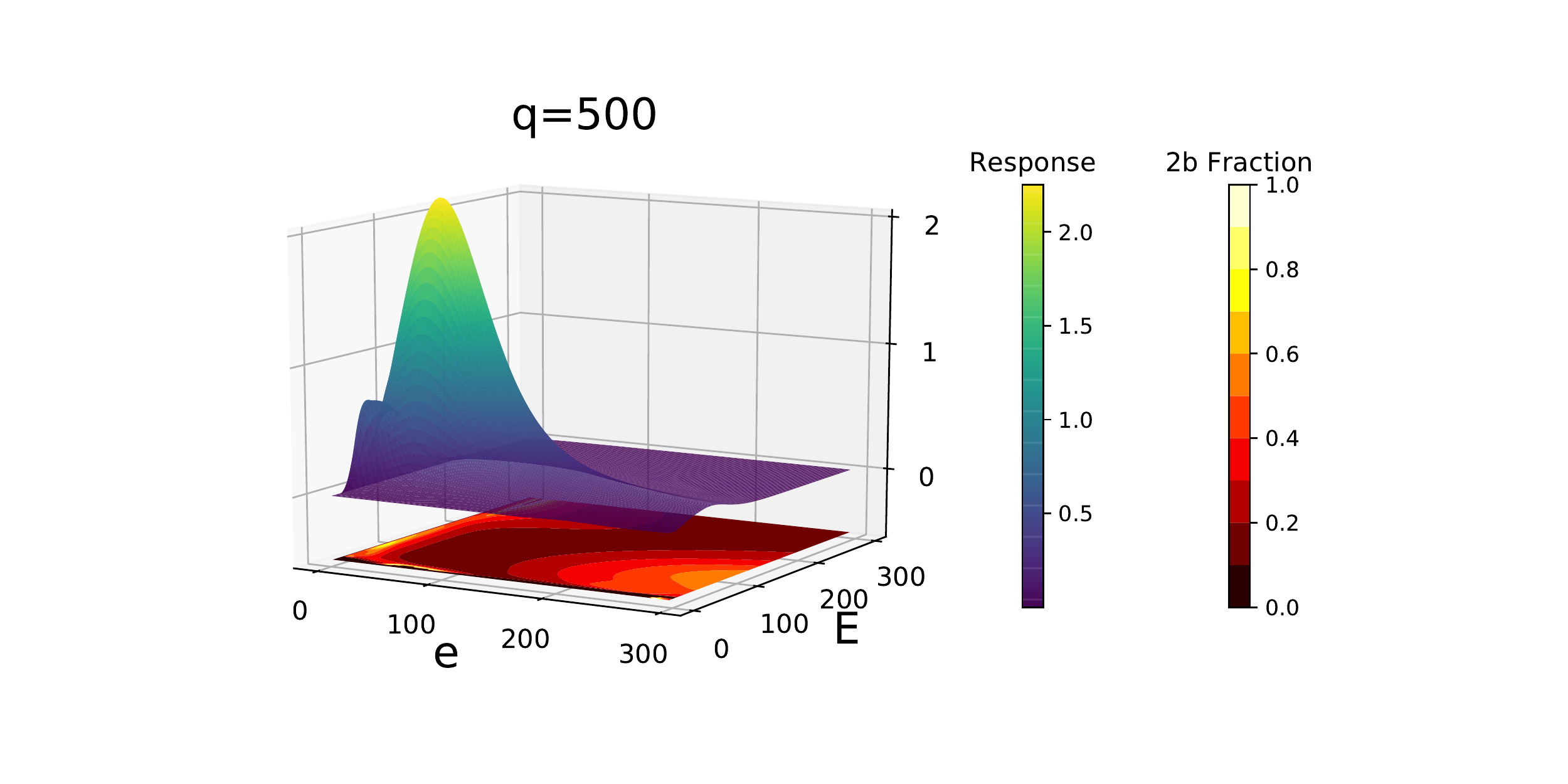}
\caption{Transverse response density at $q = 500$ MeV/c.
         The 3D plot shows the response density as a function of relative $e$ and center-of-mass $E_{\rm cm}$ energies.  
         The contour plot below shows the fraction
	 of the response coming from terms including two-nucleon currents.}
\label{fig:3donevstwo}
\end{figure}

We can further examine the relative contributions of one- and two-nucleon
currents at the vertex for different combination of $e$ and $E_{\rm cm}$.
In Fig.~\ref{fig:3donevstwo} we again show the transverse response density
at $q\,$=$\,500$ MeV/c as a function of $e$ and $E_{\rm cm}$. As expected the response
is significant out to high relative energies $e$ because of the two-nucleon
currents and correlations.  The figure also shows a contour plot of the
fractional component of the response densities which include two-nucleon
currents (either interference or pure two-body terms).  The fraction is
small at low relative energies, but increases to approximately fifty percent
at high relative energies in the pair at the vertex.  This is what we
expect based upon the analytical arguments presented in Sec.~\ref{sec:twobodyphysics}, 
of course the calculation includes the full set of two-nucleon currents, 
not only the pion seagull piece.
The average contribution of the two-nucleon currents is roughly thirty
percent, as demonstrated by the sum rules.  It is even higher in the
regime of large relative energies, or back-to-back pairs.

\section{Explicit Final States and Back-to-Back Nucleons}
\label{sec:exclusiveresults}

The additional information about the states immediately after the electromagnetic vertex
at the two-nucleon level can be used to gain insight on the cross section for explicit final
states.  In light nuclei the information at the vertex will be closely
correlated with the observed final state. In a larger nucleus, event generators will be required to go from
the vertex-level description provided by the STA to the full
final state interaction.  The event generators provide an essentially classical description
of the final state interactions after the two-nucleon vertex.   As we have discussed,
quantum interference between initial state interactions and two-nucleon currents is
important to produce the correct vertex environment.  However, subsequent evolution
is expected to be largely classical.  Further tests of this method may be possible
using quantum computers~\cite{Roggero:2018hrn}, which can at least in principle perform
the full quantum evolution of the final states.

Many experiments have been performed looking at back-to-back kinematics
for proton-neutron versus proton-proton and neutron-neutron pairs~\cite{Hen614,Arrington:2011xs,Fomin:2017ydn}.
Given the experimental interest in these special kinematical configurations,
it is interesting to compare the STA response coming from these different types
of pairs initially at large relative momentum and small center-of-mass momentum.  These pairs
in the back-to-back kinematics (that is, pairs with initial center-of-mass momentum equal to zero)
can be isolated in the response densities by choosing
a pair with final center-of-mass momentum $P$ close to the momentum transfer $q$, with large relative
momentum in the final state.

\begin{figure}[!htbp]
    \begin{center}
    \begin{minipage}{0.45\textwidth}
        \centering
        \includegraphics[height=2.2in]{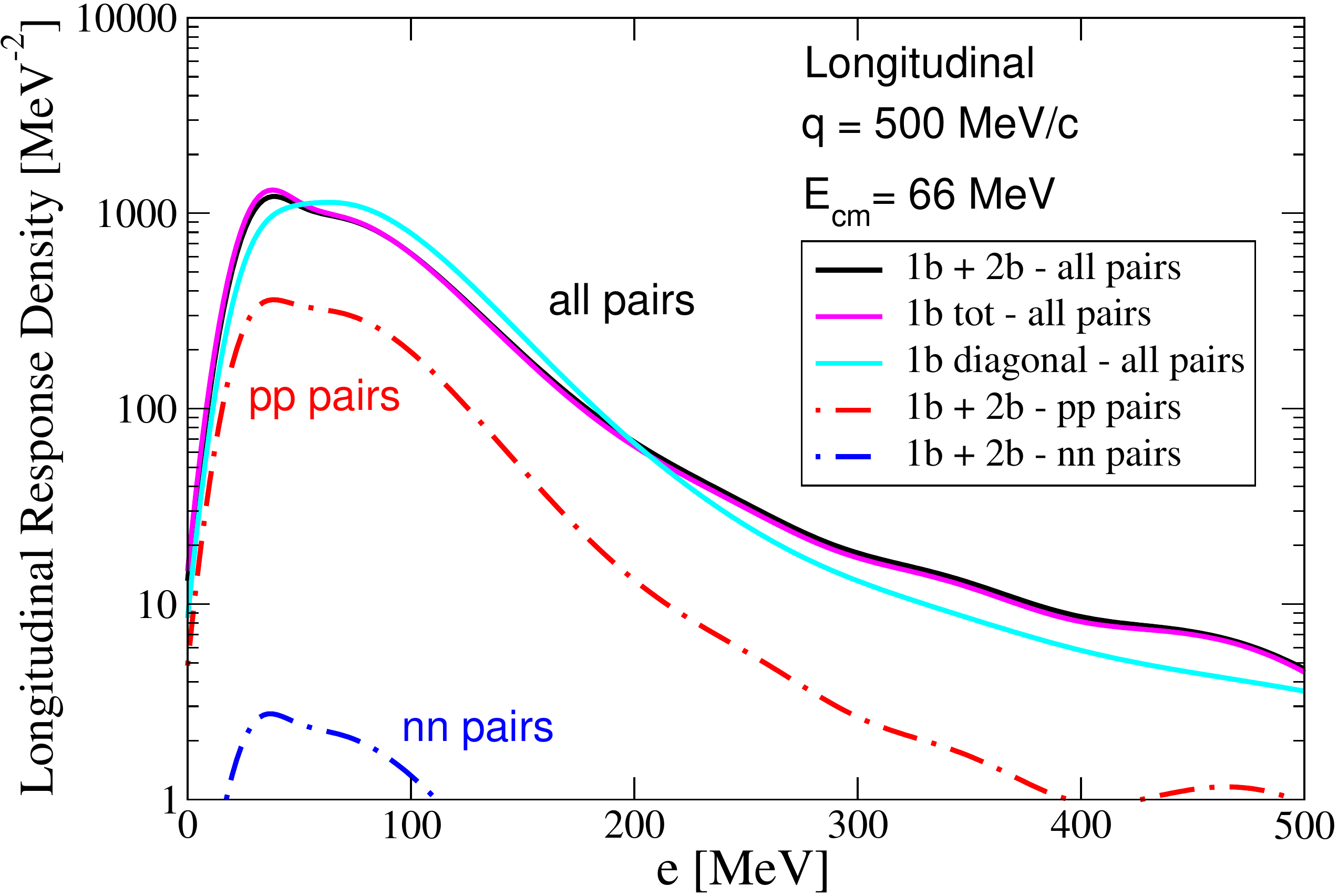}
    \end{minipage}
    \hspace*{0.6cm}
    \begin{minipage}{0.45\textwidth}
        \centering
        \includegraphics[height=2.2in]{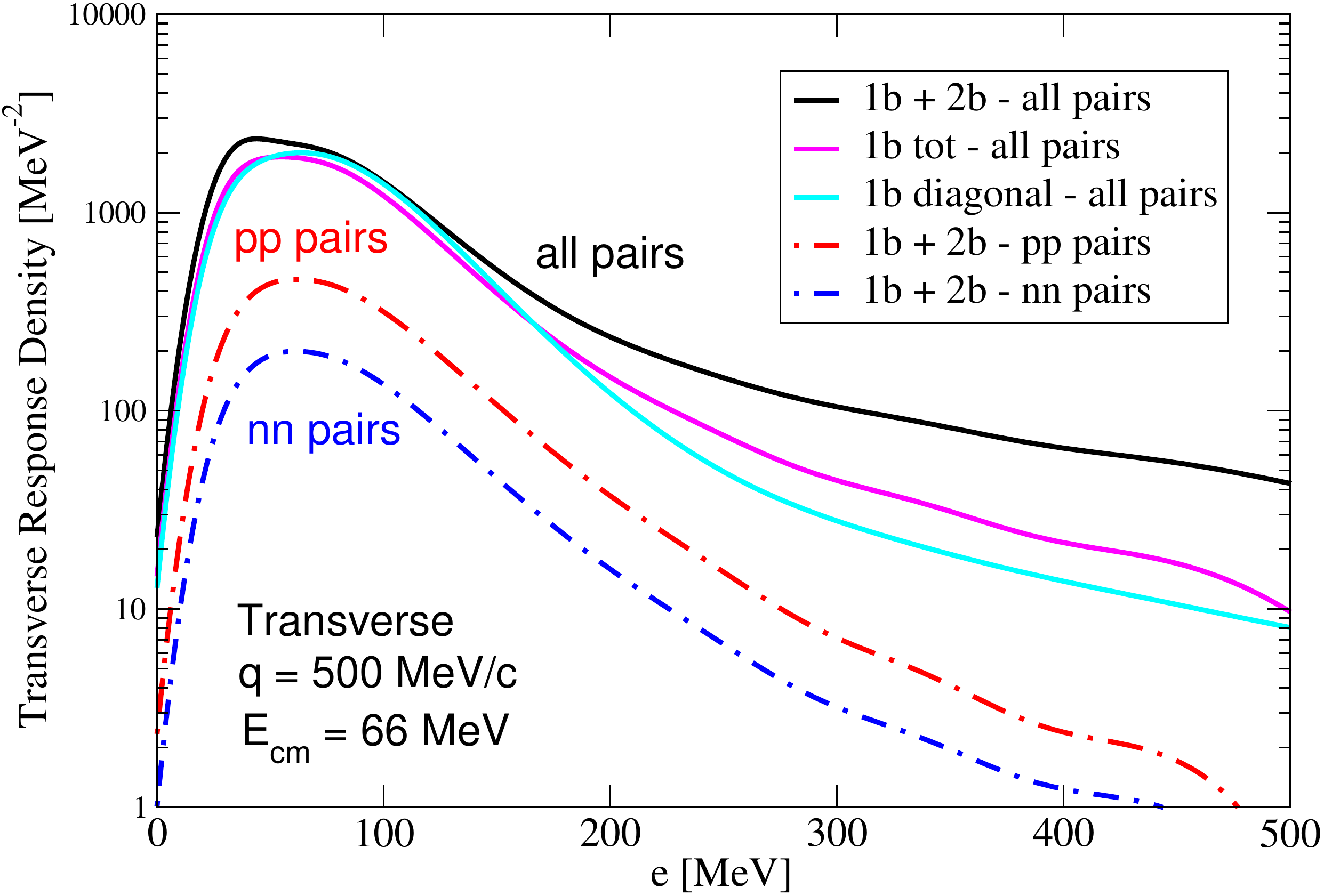}
    \end{minipage}
    \end{center}

    \caption{ Contributions to response densities at $q=500$ MeV/c and 
    final center of mass energy $E_{\rm cm}= q^2/(4m)$. Contributions of one-body diagonal terms are shown (cyan solid line),
    along with the total one-body currents given by diagonal plus off-diagonal contributions (magenta solid line). 
    Full (one- plus two-body currents) results are also shown for both total (solid black line) 
    and contributions from $nn$ (blue dashed line) and $pp$ pairs (red dashed line).   
    Left panel: Longitudinal.  Right panel: Transverse.}
    \label{fig:backtobacklt500}
\end{figure}

\begin{figure}[!htbp]
    \begin{center}
    \begin{minipage}{.45\textwidth}
        \centering
        \includegraphics[height=2.2in]{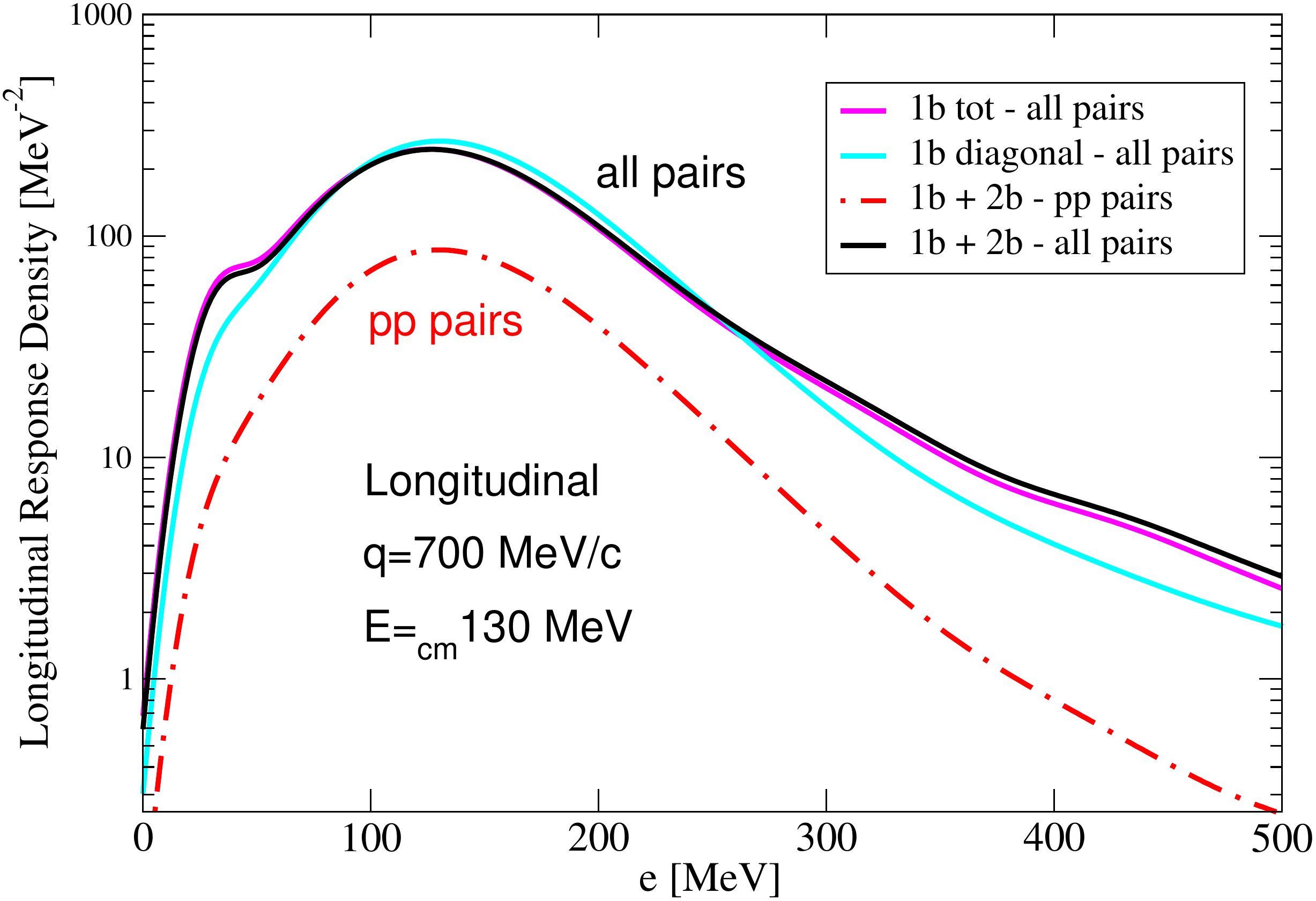}
    \end{minipage}
    \hspace{0.5cm}
    \begin{minipage}{0.45\textwidth}
        \centering
        \includegraphics[height=2.2in]{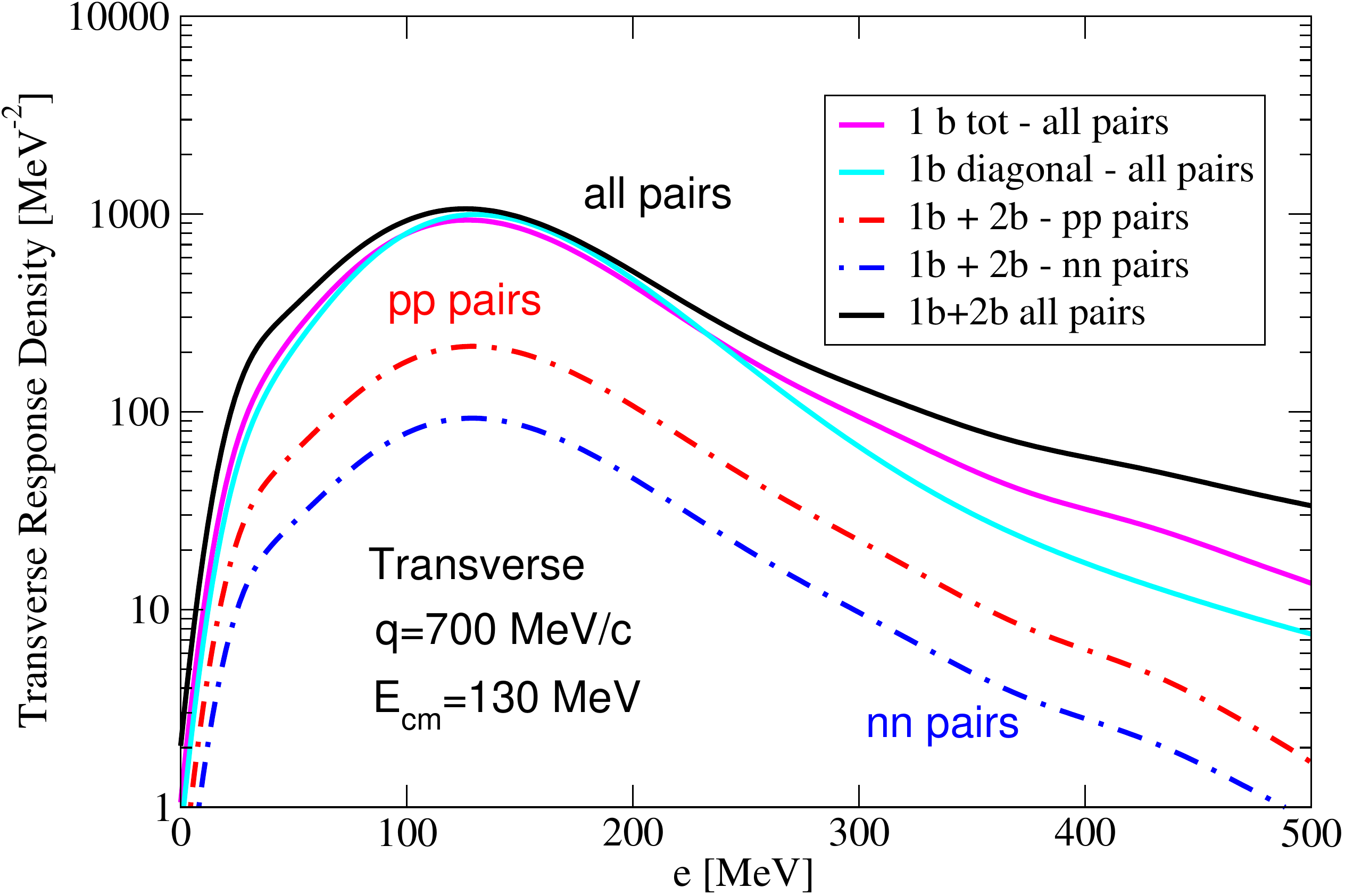}
    \end{minipage}
    \end{center}

    \caption{ Same as Fig.~\ref{fig:backtobacklt500} but at $q = 700$ MeV/c.}
    \label{fig:backtobacklt700}
\end{figure}

Figures~\ref{fig:backtobacklt500} and~\ref{fig:backtobacklt700} show the
response densities at fixed energy $E_{\rm cm} \sim  P^2/(4m) = q^2/(4m)$, which is the final state 
center-of-mass energy for an initial pair with total momentum zero, as function
of the relative energy of the pair. The regime of large
back-to-back momentum is above $e\,$=$\,250$ MeV which corresponds to the final
pair relative momentum of $\sim 2.5$ fm$^{-1}$ and above.
On the left panels the longitudinal responses are shown, including the full 
response, the one-body diagonal and one-body total ({\it i.e.}, diagonal plus 
off-diagonal one-body terms) response, and the response from $pp$ pairs.  
Note there are essentially no $nn$ pairs because
the charge form factor of the neutron is very small. In the longitudinal
channel almost all the response comes from the one-body currents, as expected.
Nevertheless, there is a large contribution of back-to-back $np$ pairs
because there are four $np$ pairs and only one $pp$ pair in the alpha particle.
We note that the $pp$ pairs always have a finite contribution because
of coherent interference in the contributions from different nucleons; the
latter fill in the node in the $pp$ contributions arising from
the $pp$ momentum distribution~\cite{Schiavilla:2006xx} at zero total momentum.

On the right hand side of the figures the transverse response density is
shown for the same kinematics.  At low relative energies the two-nucleon 
currents are not making large contributions, as shown by comparing the full 
results (black line) to the one-body total results (magenta line).  However,
at high relative energy the full result (black line) is substantially 
larger than the full one-body currents calculations (magenta line).
The back-to-back momentum distributions of $np$ pairs are known to dominate
over $pp$ or $nn$ pairs at high relative momenta~\cite{Schiavilla:2006xx}. However, 
this is not the complete picture. The $np$ pairs receive a substantial contribution
from two-nucleon currents, as expected based upon the arguments above.
These two-nucleon currents are almost entirely in the $np$ pairs, and 
increase the response by roughly a factor of $\sim2$ around at $e\,$=$\,300$ MeV.

We note that purely hadronic experiments~\cite{Tang:2003} also show an enhancement of $np$ versus
$nn$ or $pp$ pairs. The relative momentum distribution is relevant and is 
much enhanced in the $np$ channel. Further studies of final state interaction
effects in these experiments are warranted to make a detailed comparison with
experimental results.

The $nn$- and $pp$-pair contributions are almost all from one-body currents. The responses of $nn$ and $pp$ pairs differ, since
the magnetic moments of the proton and neutron $\mu_p$ and $\mu_n$
are different; indeed, the $pp$-to-$nn$ response ratio
scales roughly as $(\mu_p/\mu_n)^2$.

\begin{figure}[!htbp]
    \begin{center}
    \begin{minipage}{0.45\textwidth}
        \centering
        \includegraphics[height=2.in]{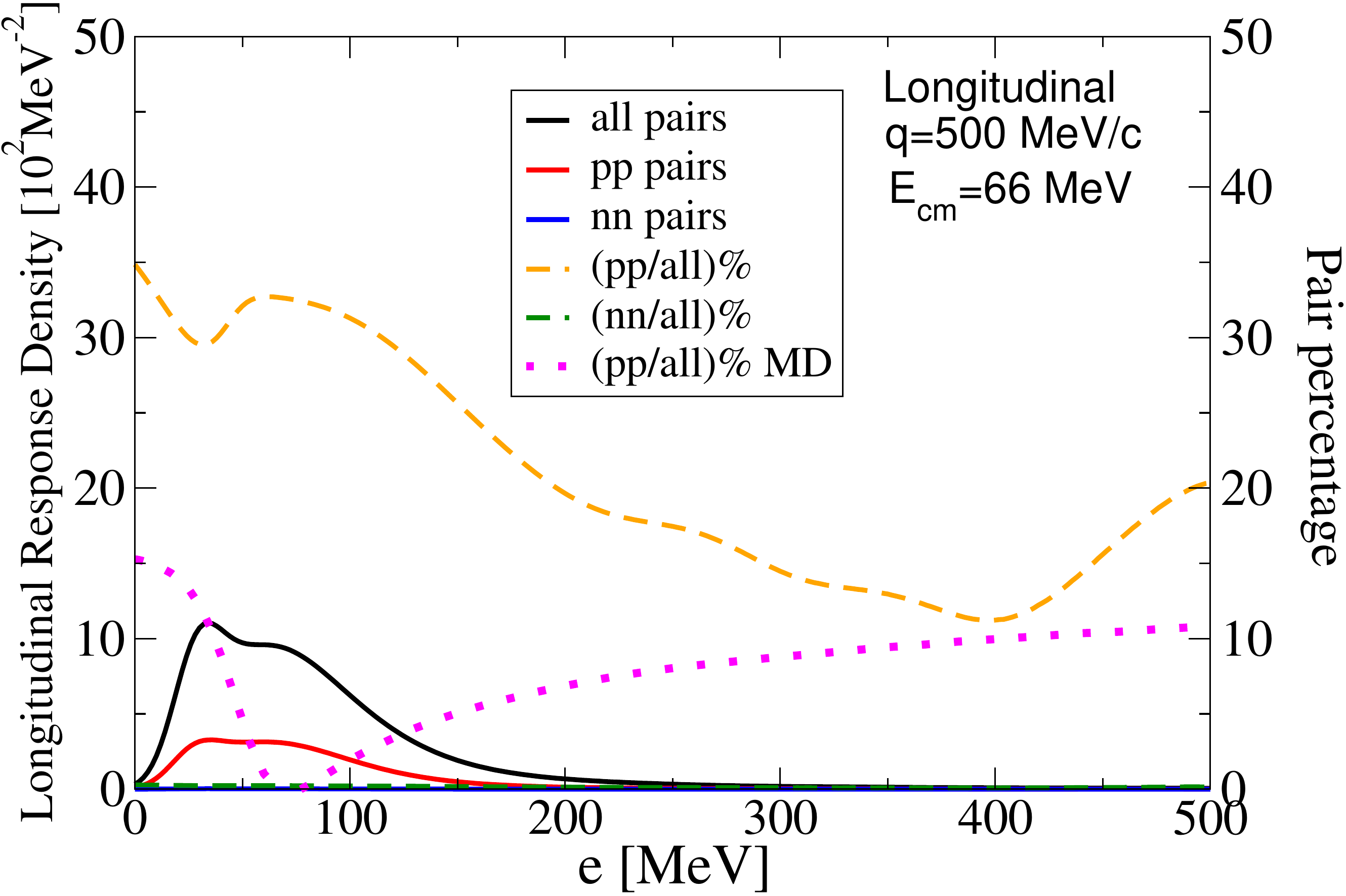}
    \end{minipage}
    \hspace*{0.6cm}
    \begin{minipage}{0.45\textwidth}
        \centering
        \includegraphics[height=2.in]{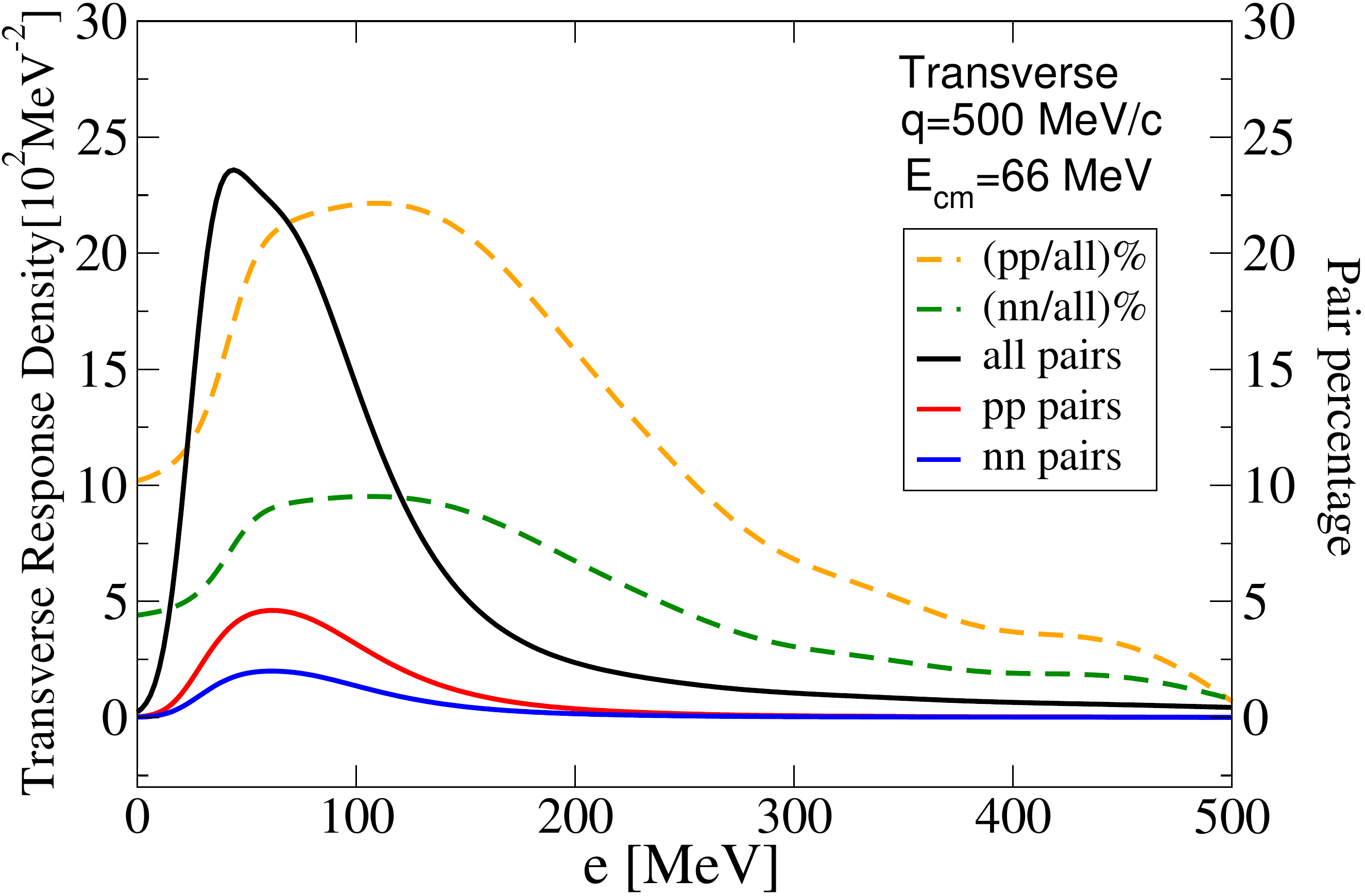}
    \end{minipage}
    \end{center}

    \caption{ Contributions to response densities at 
$q=500$ MeV/c and  final center of mass energy $E_{\rm cm}= q^2/(4m)$. Total contributions from
all pairs (black solid  line) and from from $nn$ (blue solid line) and $pp$ (red solid line) pairs
are shown along with the percentage of $pp$ (dashed orange line) and $nn$ (dashed dark green line)
contributions over the all pairs contribution.  
Left: Longitudinal. Here, the magenta dotted line represents the $pp$ pair percentage from
two-nucleon momentum distributions---MD in the figure---from Ref.~\cite{Wiringa:2013ala}.
Right: Transverse.}
    \label{fig:percentage500}
\end{figure}

In Fig.~\ref{fig:percentage500}, we show the percentage of response density in 
the back-to-back configuration at $q\,$=$\,500$ MeV/c due to scattering from $pp$ and $nn$ pairs. 
This is the ratio of the response due to scattering from $pp$ ($nn$) pairs over the full response.
In the longitudinal response (left panel), relevant to scattering in the forward direction, 
at high relative energies $e \sim 300$ MeV and above, 
the percentage due to $pp$ pairs is of the order of $\sim 15\%$, 
while neutron pairs contributions are negligible due to the small electromagnetic
nucleonic form factor. In the transverse response, at $e \sim 300$ MeV, we see a 
$\sim 10\%$ contribution from $pp$ pairs versus a $\sim 5\%$ contribution from 
$nn$ pairs, again primarily due to the different proton and neutron
magnetic moments.  In the left panel of Fig.~\ref{fig:percentage500},
we show for comparison the ratio of the (two-body) $pp$ momentum distribution
over the total two-body momentum distribution from Ref.~\cite{Wiringa:2013ala}.
This is given in the figure by the magenta dotted line.  As discussed above, 
the pair percentage estimated from the two-body momentum distributions exhibits
a deep which is filled in when interference between one-body and one- and
two-body currents are accounted for (orange
dashed line in the figure).

\section{Summary}
\label{sec:summary}
\begin{figure}[htbp]
\centering
\includegraphics[width=\linewidth]{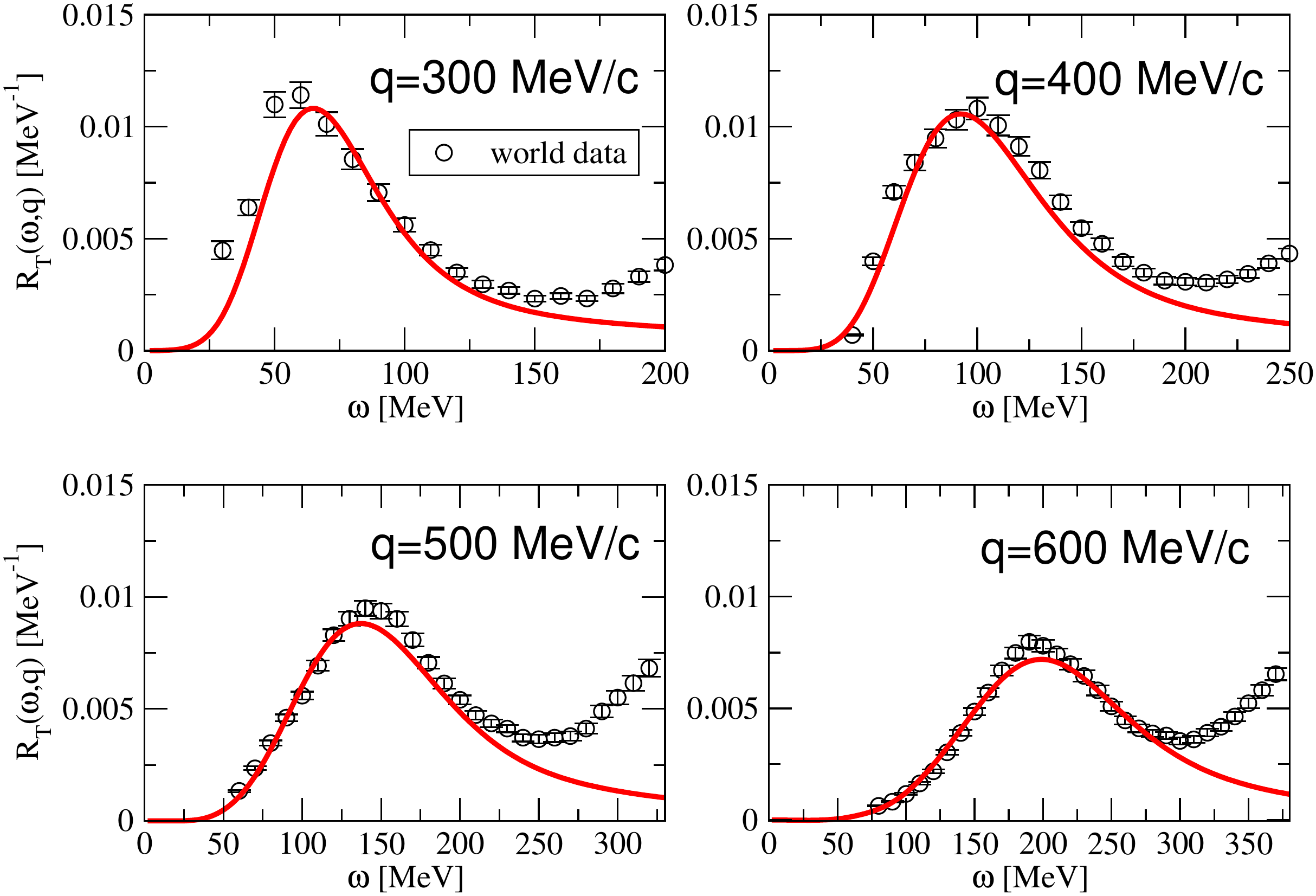}
\caption{Transverse response functions calculated with the STA at $q \,$=$\, 300$--$600$ MeV/c
are compared with those obtained from analysis of the world data~\cite{Carlson:2001mp}.}
\label{fig:worlddata}
\end{figure}

In this work we introduced the short-time-approximation which, when combined
with Quantum Monte Carlo computational methods, allows one to evaluate
nuclear response functions and response densities. We showed that calculations
within the STA accurately reproduce the quasi-elastic response of light
nuclei at momentum transfers near and above the Fermi momentum.
In this regime the STA- and GFMC-calculated response functions are in
very good agreement with each other.  A comparison of the STA transverse
response functions with those extracted from an analysis of the world data~\cite{Carlson:2001mp} 
and shown in Fig.~\ref{fig:worlddata} also indicates there is
excellent agreement between STA-theory and experiment 
for momentum transfers in the range $q\,$=$\,300$--600 MeV/c.

The STA incorporates all the important two-nucleon physics systematically, 
including ground-state correlations, two-nucleon currents, and
final state interactions.  All of these elements quite important, particularly
the interference between OPE correlations and currents. 
The STA goes beyond the
spectral function approach in explicitly taking into account the specific 
electroweak two-nucleon current operators and Pauli blocking between 
the struck and spectator nucleons.  The cost is that it must be evaluated explicitly
in the ground state for each momentum transfer $q$ and each transition
current operator.

Additionally, the STA provides information about pairs of 
nucleons at the interaction vertex.
This can be very valuable when trying to understand more exclusive processes
like back-to-back nucleons that can be measured experimentally. It is also 
important in neutrino physics, where analyses of specific final states are
used to gain information on the initial neutrino energy, a crucial input
in neutrino oscillation analyses.  For large nuclei this information about
the vertex will have to be augmented by semi-classical event generators.

The STA is amenable to many improvements, particularly in the 
higher-energy regime. Since it factorizes the response into a
two-nucleon component and a spectator nucleus, one can more easily incorporate 
relativistic kinematics and currents, pion production, and resonance production.  
Treating such effects at the two-nucleon level is vastly easier than calculating the same processes
in a full $A$-nucleon treatment. We expect that interference effects, for example from different
processes leading to pion production, may be important there as well.

\vspace{0.5cm}
We wish to thank  
S.\ Bacca for sharing her LIT results,
D.\ Lonardoni for his calculations used to 
obtain nearest-neighbor distributions in nuclear matter,
A.\ Lovato for sharing his GFMC results,
I.\ Sick for providing us with the experimental data on the $^4$He response functions,
and J.L.\ Barrow, M.\ Betancourt, S.\ Gardiner,  R.\ Castillo-Fernandez, and G.\ Perdue for 
useful discussions at different stages of this work.
The work of SP has been supported by the FNAL Intensity Frontier Fellowship and
by the U.S.~Department of Energy funds through the Neutrino Theory Network and through
the FRIB Theory Alliance award DE-SC0013617. 
The work of SP, JC and  SG has been supported by the NUclear
Computational Low-Energy Initiative (NUCLEI) SciDAC project.  This research
is also supported by the U.S.~Department of Energy, Office of Science,  Office of
Nuclear Physics, under contracts DE-AC05-06OR23177 (RS),
DE-AC02-06CH11357 (RBW), and DE-AC52-06NA25396 and 
Los Alamos LDRD program (JC~and SG).
The work of SG was also supported the DOE Early Career Research Program.
This research used resources allocated by the Argonne Leadership Computing 
Facility at Argonne National Laboratory, and by the Argonne's Laboratory Computing Resource Center.
It was also supported by the US Department of Energy through the Los
Alamos National Laboratory and used resources provided by the Los Alamos National
Laboratory Institutional Computing Program. Los Alamos National Laboratory is operated
by Triad National Security, LLC, for
the National Nuclear Security Administration of U.S. Department of Energy
(Contract No.\ 89233218CNA000001).
We also used resources provided by NERSC, which is supported by the US
DOE under Contract DE-AC02-05CH11231.

\appendix
\section{Two-nucleon currents and correlations}
\label{app:inter}

In this Appendix, we work in more detail the algebra used to 
support the discussion on the relevance of two-body physics reported in 
Sec.~\ref{sec:twobodyphysics}. Throughout the course of this
Appendix we will refer to the diagrams illustrated in the panels of Fig.~\ref{fig:diagrams}.

\subsection{Mean Fields and one-body currents}
\label{sec:meanstrees}

The goal is to sketch the derivation of the current-current
amplitudes entering Eq.~(\ref{eq:sumrule}) and show that the excess 
in the electromagnetic transverse strength is explained by accounting
for two-body effects in both the nucleonic correlations and currents
of one-pion range. As discussed in Sec.~\ref{sec:twobodyphysics}, 
these amplitudes---where we retain up to two-body terms in the Hamiltonian
and currents---are represented by the diagrams illustrated in Fig.~\ref{fig:diagrams}.

We first consider the amplitude associated to the left diagram
in Fig.~\ref{fig:diagrams}. This corresponds to the mean field
contribution to the sum rules. The isovector component of the
one-body electromagnetic current operator~\cite{Carlson:1997}, 
represented in the figure by the  vertex with a photon hooking up to a nucleon line, 
is given by
\begin{equation}
\label{eq:onebody}
	{O}_1 ({\bf q}) = G_M^V (Q^2) \frac{-i}{2\,m} \, \mu_{ V} \frac{\tau_{i,z}}{2} \, 
	{\bm \sigma}_i \times {\bf q } \, , 
\end{equation}
where $\mu_{\rm V}\sim 4.7$ n.m. and $G_M^V (Q^2)$ are the nucleonic isovector 
magnetic moment and form factor, respectively.

Upon calculating the incoherent current-current term in the squared matrix element one 
obtains an amplitude, $M_a$, given by 
\begin{equation}
	| M_a |^2 =  [G_M^V(Q^2)]^2  \frac{1}{16 \, m^2} \, \mu_{ V}^2
	( {\bm \sigma}_i \times {\bf q} ) \cdot
	( {\bm \sigma}_i \times {\bf q} ) \, ,
\end{equation}
which reduces to 
\begin{equation}
\label{eq:mmeanfield}
	| M_a |^2 =  [ G_M^V (Q^2) ]^2 \, \frac{1}{8\, m^2} \, \mu_V^2 \, q^2 \, ,
\end{equation}
when keeping only diagonal terms in the squared matrix element.
With this approximation, the sum rule is given by the quantity above
summed over the nucleons, here including the dominant isovector terms.

\subsection{High-momentum nucleons and one-body currents}
\label{sec:highonebody}

The second diagram in Fig.~\ref{fig:diagrams} illustrates the contribution of 
high-momentum nucleons---that is nucleons with higher momenta induced  by 
the one-pion-exchange correlations---with one-body currents. In this case,
using the notation introduced in Eq.~(\ref{eq:Lippmann}), we can schematically 
write the correlated two-nucleon wave function, $\psi_\pi$, as
\begin{eqnarray}
	\psi_\pi (k) &\sim&  - \frac{ v_\pi }{E(k)} |\ 0 \rangle 
	             = \frac{-1}{E(k)}\,\widetilde{ v}_\pi(k) 
		  \    {\bm \sigma_i} \cdot {\bf k}
		  \    {\bm \sigma_j} \cdot {\bf k} \ {\bm \tau}_i \cdot {\bm \tau}_j
	| \ 0 \rangle \, ,
\end{eqnarray}
where we defined 
\begin{equation}
\widetilde{v}_\pi (k)  = - \frac{ f_{ {\pi} NN}^2}{3\, m_\pi^2} \frac{1}{ k^2 + m_\pi^2} \, , 
\end{equation}
with $f_{ {\pi} NN}$ and $m_\pi$ being the pion-nucleon coupling constant and 
the pion mass, respectively. 
In the equations above, $| \ 0 \rangle$ is the mean-field ground state which  
contains low-momentum nucleons.  Also, the one-pion exchange interactions,
$\tilde{v}_\pi(k)$, gives momentum ${\bf k}$ to nucleon $j$ and $-{\bf k}$ to nucleon $i$.

We now consider the incoherent sum over single-nucleon currents
arising from the high-momentum ground state components induced by
pion exchange.  This contribution is given by
\begin{eqnarray}
	|M_{\pi}|^2 & = & 
	\langle 0 | \frac{ - {\widetilde v}_\pi (k)}{E(k)}
	{\bm \tau}_i \cdot {\bm \tau}_j
	{\bm \sigma}_i \cdot {\bf k} {\bm \sigma}_j \cdot {\bf k} 
	\nonumber \\
	& &
	\left(\frac{\mu_V}{2m}\right)^2 G_M^V (Q^2)^2 \frac{1}{4}
        {\bm \tau}_{i,z}^\dagger {\bm \tau}_{i,z}  \ \ 
	({\bm \sigma}_i \times {\bf q}) \cdot ({\bm \sigma}_i \times {\bf q} )
	\nonumber \\ & &
	\ {\bm \sigma}_i \cdot {\bf k}\  {\bm \sigma}_j \cdot {\bf k}
	\ {\bm \tau}_i \cdot {\bm \tau}_j \ 
	 \frac{ - {\widetilde v}_\pi (k)}{E(k)} | 0 \rangle \nonumber \\
	 & = & 
	\langle 0 | \left[\frac{ - {\widetilde v}_\pi (k)}{E(k)}\right]^2
	\left[ \left(\frac{\mu_V}{2m}\right)^2 \  G_M^V (Q^2)^2 \ \frac{1}{2} \right]
	(3 - 2 {\bm \tau}_i \cdot {\bm \tau}_j  )
	k^4 q^2 | 0 \rangle
\end{eqnarray}

Here we have assumed that the initial momenta in $| 0 \rangle$ is small
and that the momentum transfer $q$ is large, larger than $k_F$
(see middle panel in Fig.~\ref{fig:diagrams}). 
If we further assume the high-momentum pairs are primarily $S=1, T=0$ pairs,
average over the directions of q, 
and estimate the  energy denominator as $E(k) \sim k^2/m$
we get
\begin{equation}
	M_{\pi}^2  =  \frac{9}{8} \ 
	 \widetilde{v}_\pi (k)^2  \ q^2  \ \mu_V^2  \ G_M^V (Q^2)^2
\end{equation}

Note that these high-momentum nucleons spread the response out in energy, 
the initial
momentum can be either parallel or anti-parallel to q so the response
is broadened.  The same happens with the interference term described below.

\subsection{ Interference of One- and Two-Body Currents}

The third diagram in Fig.~\ref{fig:diagrams} illustrates the interference 
between one- and two-body currents acting on correlated two-nucleon states.
We again consider the pion-correlation contribution to the ground state 
wave function in first-order perturbation theory, as outlined in Eq.~(\ref{eq:Lippmann}),
and assume that the one-pion-exchange interaction, $v_\pi$, gives momentum 
${\bf k}$ to nucleon $j$ and  $- {\bf k}$ to nucleon $i$.
Further, for this example assume 
the only important interference is with the current hitting on 
nucleon $i$ with the simultaneous exchange of a pion with nucleon $j$
(see right panel in Fig.~\ref{fig:diagrams}). This current contribution of 
one pion range is called seagull term~\cite{Carlson:1997,Carlson:1994zz}.

As in the Sec.~\ref{sec:meanstrees}, the one-body current vertex acting on 
nucleon $i$  then gives a factor
\begin{equation}
	{O}_1 ({\bm q}) = G_M^V (Q^2) \frac{-i}{2\,m} \, \mu_{ V} \frac{\tau_{i,z}}{2} \, 
	{\bm \sigma}_i \times {\bf q } \, , 
\end{equation}
while the product 
$O_2^\dagger (q) O_1 (q)$ is

\begin{eqnarray}
	O_2^\dagger O_1 \  & = \  &
	(-3i) \ {\bm \tau}_i \times  {\bm \tau_j} \cdot {\hat z}  \tau_{i,z}\ 
	G_E^V (Q^2) G_M^V (Q^2) \ \frac{-i}{2m} \  \frac{ \mu_V}{2} 
	\ {\widetilde v}_\pi (k)  \  {\bm \sigma}_j \cdot {\bf k} \ 
	{\bm \sigma}_i  \cdot ({\bm \sigma}_i \times {\bf q})
	\nonumber \\
	& = \ &   \frac{-3}{2}
	\left[ {\bm \tau}_i \cdot {\bm \tau}_j - \tau_{i,z} \tau_{j,z} \right]
	G_E^V (Q^2)  \  G_M^V (Q^2) \  \frac{-i}{2m} \ \mu_V \ {\widetilde v}_\pi (k)  \ 
	{\bm \sigma}_j \cdot {\bf k} \ 
	{\bm \sigma}_i  \cdot ({\bm \sigma}_i \times {\bf q})
	\nonumber \\
	& = &  \frac{-3 }{2} \ 
	\left[ {\bm \tau}_i \cdot {\bm \tau}_j - \tau_{i,z} \tau_{j,z} \right] \ 
	G_E^V (Q^2)  \  G_M^V (Q^2) \  \frac{1}{2m} \ \mu_V \ {\widetilde v}_\pi (k)\ 
	{\bm \sigma}_j \cdot {\bf k} \ 
	{\bm \sigma}_i \cdot {\bf q} 
\end{eqnarray}

Then the quantity  $O_2^\dagger O_1 \psi_\pi (k) $ is
\begin{eqnarray}
	O_2^\dagger O_1 \psi_\pi (k) \  & = \  &
	\frac{-3 }{2} \ 
	\left[ {\bm \tau}_i \cdot {\bm \tau}_j - \tau_{i,z} \tau_{j,z} \right] \ 
	G_E^V (Q^2)  \  G_M^V (Q^2) \  \frac{\mu_V}{2m} \ {\widetilde v}_\pi (k) \  
	{\bm \sigma}_j \cdot {\bf k} \ 
	{\bm \sigma}_i \cdot {\bf q} 
	\nonumber \\
	&\times & \frac{- {\widetilde v}_\pi(k)}{E (k)} \ {\bm \tau}_i \cdot {\bm \tau}_j
	\ { \bm \sigma}_i \cdot {\bf k}\  { \bm \sigma}_j \cdot {\bf k}\, | 0 \rangle
	\nonumber \\
	& = & 
	\frac{ 3 }{2} 
	\left[ ({\bm \tau}_i \cdot {\bm \tau}_j)^2 - 
	({\bm \tau}_i \cdot {\bm \tau}_j) \tau_{i,z} \tau_{j,z} \right] 
	G_E^V (Q^2)   G_M^V (Q^2) \  \frac{\mu_V}{2m}  \frac{{\widetilde v}_\pi (k)^2}{E (k)}
	\nonumber \\ & \times& 
	 k^2 \left[ {\bf k} \cdot {\bf q} + i {\bm \sigma}_i \cdot ( {\bf k} \times {\bf q}) \right] \,| 0 \rangle
\end{eqnarray}

Only the part of ${\bf k}$ perpendicular to ${\bf q}$ enters the last
term, and it will average to zero. The isospin factor is $6$ in T=0 pairs.
If we again put $E (k) = k^2/m$ the matrix element for T=0 pairs
simplifies to
\begin{equation}
	|M_{\rm int}|^2=\langle 0 | O_2^\dagger O_1 \psi_\pi (k) \rangle
	= \frac{9}{2}  G_E^V (Q^2) G_M^V (Q^2)  \ {\widetilde v}_\pi (k)^2 \ \mu_V \ {\bf k} \cdot {\bf q}
\end{equation}

Constructive interference will occur for $\bf k$ parallel to $\bf q$,
while for $\bf k$ antiparallel to $\bf q$ the terms will have opposite
signs, which implies $|M_{\rm int}|^2\propto G_E^V (Q^2) G_M^V (Q^2)  \ {\widetilde v}_\pi (k)^2 \ \mu_V \,q^2$.
In this appendix we have considered only the seagull term with
the combined strong and EW vertex at one nucleon, and also one time ordering.

If we assume the dominant contributions come from k in the same direction
to q with a 'typical' momenta of $k = q/2 $ to $k=q$ we get a 
ratio of contributions of the interference term to the high-momentum
component of
\begin{equation}
	\frac{ \langle 0 | O_2^\dagger O_1 \psi_\pi (k) \rangle}
	{ \langle \psi_\pi (k) | O_1^\dagger O_1 | \psi_\pi (k) \rangle}
	\sim ( 1/2 \ {\rm to} \ 1) 
	\ \frac{4}{\mu_V} \ \frac{G_E^V (Q^2) }{G_M^V (Q^2)} \  
	\sim ( 1/2 \ {\rm to \ 1})
\end{equation}
which says that two-body physics both in the correlations and current provide
corrections of the same order that add up constructively with the 
mean-filed low-momentum amplitude of Eq.~(\ref{eq:mmeanfield}).

Note that the contribution of the other 'seagull' diagram gives zero
at ${\bf k}\,$=$\, {\bf q}$ (just as this diagram gives zero at ${\bf k} = 0$), 
while at ${\bf k} = {\bf q}/2$ the two contributions
are equal.  The two time orderings of these diagrams give 
equal contribution, as do
the two incoherent high-momentum nucleons arising from
single-nucleon currents.  Taking all this into account we expect the
interference between one- and two-nucleon terms to be similar in
magnitude to the incoherent scattering from high-momentum nucleons.

\bibliography{biblio}

\end{document}